\documentclass[final]{siamart190516}
\usepackage[letterpaper, margin=1in]{geometry}
\usepackage{lipsum}
\usepackage{amsfonts}
\usepackage{graphicx}
\usepackage{epstopdf}
\usepackage{mathtools}
\usepackage[caption=false]{subfig}
\usepackage{algorithmic}
\DeclareGraphicsExtensions{.eps}

\newsiamremark{remark}{Remark}
\newsiamremark{hypothesis}{Hypothesis}
\crefname{hypothesis}{Hypothesis}{Hypotheses}
\newsiamthm{claim}{Claim}

\usepackage{lineno,hyperref}
\usepackage{amssymb,amsmath,float,mathabx}

\modulolinenumbers[5]
\allowdisplaybreaks
\usepackage{etoolbox} 
\newcommand*\linenomathpatch[1]{%
\cspreto{#1}{\linenomath}%
\cspreto{#1*}{\linenomath}%
\cspreto{end#1}{\endlinenomath}%
\cspreto{end#1*}{\endlinenomath}%
}
\newcommand*\linenomathpatchAMS[1]{%
\cspreto{#1}{\linenomathAMS}%
\cspreto{#1*}{\linenomathAMS}%
\csappto{end#1}{\endlinenomath}%
\csappto{end#1*}{\endlinenomath}%
}

\expandafter\ifx\linenomath\linenomathWithnumbers
\let\linenomathAMS\linenomathWithnumbers
\patchcmd\linenomathAMS{\advance\postdisplaypenalty\linenopenalty}{}{}{}
\else
\let\linenomathAMS\linenomathNonumbers
\fi

\linenomathpatch{equation}
\linenomathpatchAMS{gather}
\linenomathpatchAMS{multline}
\linenomathpatchAMS{align}
\linenomathpatchAMS{alignat}
\linenomathpatchAMS{flalign}

\newcommand{\dint}{\displaystyle \int}

\definecolor{brightlavender}{rgb}{0.75, 0.58, 0.89}

\newcommand{\da}{\Delta \alpha}
\newcommand{\dt}{\Delta t}
\newcommand{\kn}{{k,n}}
\newcommand{\knp}{{k+1,n+1}}
\newcommand{\kp}{{k+1}}
\usepackage[normalem]{ulem}

\newcommand{\edit}[1]{\textcolor{black}{#1}}

\headers{Modeling Immunity to Malaria}{Qu, Patterson, Childs, Edholm, Ponce, Prosper, Zhao}

\title{Modeling Immunity to Malaria with an Age-Structured PDE Framework\thanks{Submitted to the editors \today. \funding{This paper is based on work supported by the American Mathematical Society Mathematics Research Communities through National Science Foundation grants \edit{1641020} and 1916439. CJE was supported by the AMS-Simons Travel Grants, which are administered by the American Mathematical Society with support from the Simons Foundation. \edit{DP} appreciates support from NSF Grant DMS-1951358. \edit{LZ was supported by the National Science Foundation grant DMS–1840265.}}\newline \hspace*{12pt}$^{\diamond}$Zhuolin Qu and Denis Patterson contributed equally to this work. }}

\author{Zhuolin Qu$^{\diamond}$\thanks{Department of Mathematics, University of Texas at San Antonio, San Antonio, TX 
  (\email{zhuolin.qu@utsa.edu}).}
  \and Denis Patterson$^{\diamond}$\thanks{High Meadows Environmental Institute, Princeton University, Princeton, NJ
  (\email{denispatterson@princeton.edu}).}
 \and Lauren \edit{M.} Childs\thanks{Department of Mathematics, Virginia Tech, Blacksburg, VA
  (\email{lchilds@vt.edu}).}
\and Christina \edit{J.} Edholm\thanks{Mathematics Department, Scripps College, Claremont, CA 
  (\email{cedholm@scrippscollege.edu}).}
  \and Joan Ponce\thanks{Semel Institute for Neuroscience and Human Behavior, University of California, Los Angeles, Los Angeles, CA 
  (\email{jponce15@asu.edu}).
  Current address: School of Mathematical Sciences and Statistics, Arizona State University, Tempe, AZ.}
   \and Olivia Prosper\thanks{Department of Mathematics, University of Tennessee, Knoxville, Knoxville, TN 
  (\email{oprosper@utk.edu}).}
  \and Lihong Zhao\thanks{Department of Applied Mathematics, University of California Merced, Merced, CA 
  (\email{lzhao33@ucmerced.edu}).}
    }

\begin{document}

\maketitle
\begin{abstract}
Malaria is one of the deadliest infectious diseases globally, causing hundreds of thousands of deaths each year. It disproportionately affects young children, with two-thirds of fatalities occurring in under-fives. Individuals acquire protection from disease through repeated exposure, and this immunity plays a crucial role in the dynamics of malaria spread. We develop a novel age-structured PDE malaria model, which couples vector-host epidemiological dynamics with immunity dynamics. Our model tracks the acquisition and loss of anti-disease immunity during transmission and its corresponding nonlinear feedback onto the transmission parameters. We derive the basic reproduction number ($\mathcal{R}_0$) as the threshold condition for the stability of disease-free equilibrium; we also interpret $\mathcal{R}_0$ probabilistically as a weighted sum of cases generated by infected individuals at different infectious stages and different ages. We parametrize our model using demographic and immunological data from sub-Saharan regions. Numerical bifurcation analysis demonstrates the existence of an endemic equilibrium, and we observe a forward bifurcation in $\mathcal{R}_0$. Our numerical simulations reproduce the heterogeneity in the age distributions of immunity profiles and infection status created by frequent exposure. Motivated by the recently approved RTS,S vaccine, we also study the impact of vaccination; our results show a reduction in severe disease among young children but a small increase in severe malaria among older children due to lower acquired immunity from delayed exposure.
\end{abstract} 
\begin{keywords}
age-structure, vector-host, immuno-epidemiological modeling, malaria, PDE
\end{keywords}
\begin{AMS}
92D30, 35Q92, 92B05
\end{AMS}

\section{Introduction}
Malaria, a parasitic disease transmitted by mosquitoes, infects hundreds of millions of people each year; the majority of cases are in sub-Saharan Africa, \edit{where the most prevalent species is \textit{Plasmodium falciparum} (\textit{P. falciparum}),} and the highest mortality burden is in young children. Immunity to malaria plays a key role in clinical outcomes and studies have shown that repeated exposure to malaria parasites promotes the development of immunity to severe disease \cite{langhorne2008immunity}. Consequently, different transmission intensities impact the rate at which humans acquire immunity to clinical disease, resulting in different distributions of protection across age in different regions \cite{dietz1974malaria,filipe2007determination}. Regions with low malaria transmission tend to have fairly equal protection against clinical disease across age, whereas the severity of symptoms in high transmission regions peaks in young children, with large proportions of the adult population asymptomatically infected \cite{cameron2015defining,smith2007standardizing}. Because asymptomatic infections may be less transmissible to mosquitoes \cite{alves2005asymptomatic}, the level of anti-disease immunity in the population feeds back into the probability of transmission, and therefore disease prevalence. This feedback loop between disease prevalence and anti-disease immunity is important to understand in the context of disease control, since control measures modify transmission intensity, and can therefore indirectly impact immunity in the population \cite{lindblade2013silent}. 
For example, malaria interventions that reduce exposure, like insecticide-treated nets and antimalarial treatments, shift the peak of severe malaria incidence to older ages \cite{ceesay2008changes, griffin2010reducing,omeara2008effect}. In regions marked by seasonal malaria transmission, the age distribution tends to be shifted towards older children because exposure to malaria parasites is less regular \cite{toure2016seasonality}.  This has important implications for how to implement life-saving interventions like intermittent preventive treatment (IPT), the periodic use of antimalarial drugs in infants and young children regardless of infection status.  Studies in northern Ghana, which experiences intense seasonal transmission, estimate that IPTi (IPT for infants) prevents 25\% of clinical malaria cases during the first year of life, compared with a 59\% reduction in clinical cases in Tanzania, where there is perennial transmission
\cite{greenwood2006intermittent}. These findings have prompted some to recommend targeting control strategies that reduce mortality, such as the use of IPT, in older children in regions with seasonal or low transmission \cite{cohee2020preventive, greenwood2006intermittent,matangila2015efficacy}.   

While mathematical modeling has been a key tool in understanding malaria dynamics for over a century \cite{macdonald1956epidemiological,ross1911prevention}, relatively few models have attempted to incorporate the dynamic feedback between acquired immunity and disease prevalence. 
The introduction of immunity to improve upon existing malaria models began in the 1970s, with the hypothesis that there were several types of acquired immunity to malaria, including loss of infectivity and loss of detectable parasite levels \cite{dietz1974malaria}. In \cite{dutertre1976study}, Dutertre incorporated acquired immunity by prolonging the time to return to susceptible in the event of re-exposure to malaria, and Elderkin et al.~\cite{elderkin1977steady} assumed that parasite-load influences resistance to infection. In the 1980s, Aron~\cite{aron1983dynamics}, in a delay-differential equation framework, included exposure-boosting immunity by assuming that immunity gradually decays over time, but will rebound if another exposure occurs within $\tau$ years. In retrospective studies involving malaria-therapy patient data, numerous works incorporated a combination of innate and adaptive immunity~\cite{molineaux1999review}.

More recently, immunity has been incorporated in models in more nuanced ways, including feedback between multiple scales through immuno-epidemiological models. Gulbudak et al.~\cite{gulbudak2017vector} developed a time-since-infection model in which the pathogen and two antibody response dynamics are tracked within the host. These within-host dynamics feed back into the epidemiological model by impacting the transmission rate of the pathogen from hosts to vectors, and host recovery rate. A nested age-structured partial differential equation (PDE) model was introduced by Cai et al.~\cite{cai2017does} to assess population-level effects of the complex within-host dynamics in an immuno-epidemiological context; they find that the impact of treatment has a larger effect in the context of lower immunity \cite{cai2017does}. Work by Vogt et al.~\cite{vogt2013impact} showed with an age-structured PDE that chronic, asymptomatic malaria infections represent an important transmission reservoir. A recent extension with time-since-vaccination  explores the role of waning and boosting of immunity on the ability to control disease, and finds that reducing the effective reproduction number below one is insufficient to guarantee loss of malaria~\cite{vogt2020impact}. Numerous other models include vaccination, but forgo immunity feedback, and, thus, are less relevant for our comparisons.

Filipe et al.~\cite{filipe2007determination} introduced an age-structured PDE model of malaria transmission tracking acquired immunity. They considered the role of partial protection from clinical disease through acquired immunity and also immunity to parasite levels through increased clearance. An extension to Filipe et al.'s work studied the trade-offs between loss of immunity acquisition due to decreased exposure and found that initial reductions following interventions may offset longer scale resurgence due to the loss of immunity~\cite{ghani2009loss}. However, the negative effect could be mitigated by a combination of vector control and vaccination strategies. Our model is motivated by Filipe et al.~\cite{filipe2007determination} and its extensions, but with several key differences. We relax the assumption of a fixed force of infection by age, allowing the force of infection to vary as immunity changes, and we allow different contributions to immunity by disease state~\cite{reluga2008backward}. Finally, we consider two types of vaccination: one motivated by the outcomes of the recently approved RTS,S vaccine, producing short-lived anti-parasite immunity \cite{cockburn2018malaria,ogeto2020malaria,penny2015timecourse}, and one modeling the potential outcomes of a blood-stage vaccine \cite{cockburn2018malaria,yman2019antibody}. Furthermore, while Filipe et al. \cite{filipe2007determination} study immunity profiles at equilibrium, \edit{we present a mathematical analysis of our model including calculation of the basic reproduction number.}

\section{Mathematical Model}
\begin{figure}[htbp!]
\centering
\includegraphics[width=0.95\textwidth,trim={0cm 0.8cm 0cm 0cm},clip]{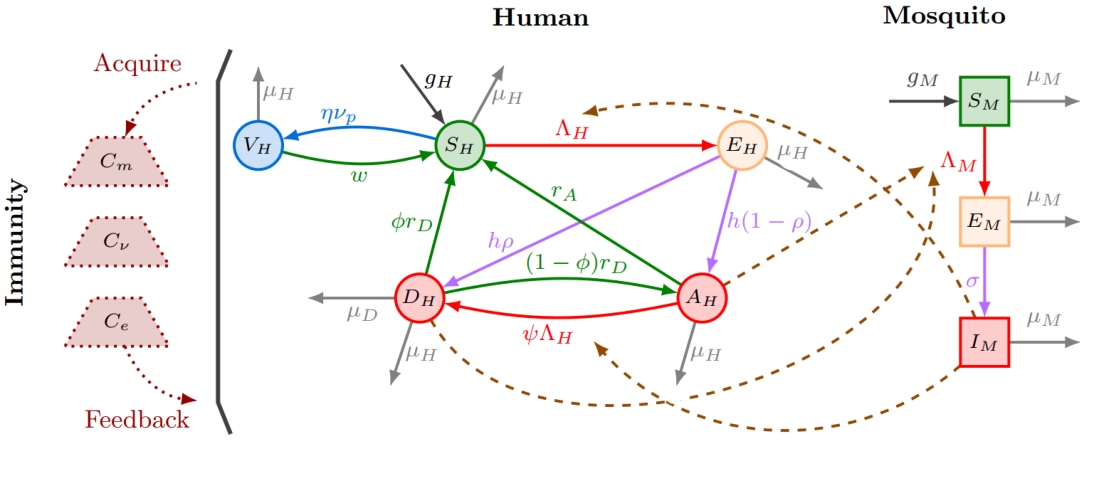}
\caption{Infection dynamics flowchart for the age-structured model. Solid arrows indicate flow of individuals, and dashed arrows indicate exposure that leads to infection. The dotted curves represent the interactions between the vector-host transmission and population immunity in humans. \label{fig:flowchart}}
\end{figure}

We propose an age-structured mixed PDE-ODE (ordinary differential equation) model to describe the \edit{\textit{P. falciparum}malaria transmission} dynamics in humans (\cref{sec:model_human}) and mosquitoes (\cref{sec:model_mosquito}). We extend this model to an immuno-epidemiological model by coupling the human-mosquito system to age-structured PDEs tracking immunity levels in the human population; human immune levels, in turn, impact the progression of the disease through nonlinear linking functions (\cref{sec:model_immune}). Our model mimics the empirically observed development of acquired immunity to malaria through repeated exposure, and loss of immunity over time via waning. A schematic diagram of the system is given in \cref{fig:flowchart}, and a summary of the state variables and parameters is given in \cref{tab:variables,tab:parameters}.

\subsection{Human Equations} \label{sec:model_human}
The human population in our model is divided into five compartments related to different stages of infection: susceptible, $S_H$; exposed, $E_H$; asymptomatic infected, $A_H$; severely diseased, $D_H$, and vaccination-protected, $V_H$. \edit{We assume individuals in $S_H$} are protected by the inherent immunity levels of the entire population, so we do not incorporate a recovered compartment. We let $P_H(\alpha,t):=S_H(\alpha,t)+E_H(\alpha,t)+A_H(\alpha,t)+D_H(\alpha,t)+V_H(\alpha,t)$ denote the number of age-$\alpha$ humans \edit{at time $t$ for $\alpha \in [0,A)$, where $A$ is the finite maximal human age, and $N_H(t):=\edit{\int_0^A} P_H(\alpha,t)\,d\alpha$ is thus the total human population.}

\edit{The susceptible human, $S_H$, is exposed upon an infectious bite from mosquitoes and enters the $E_H$ stage at the rate $\Lambda_H$. This force of infection \cref{FOI_H_revised} depends on the number of bites a person receives per time unit, $b_H$, the infectivity of infectious mosquitoes per bite, $\beta_M$, and the prevalence of infection in mosquitoes, $I_M/N_M$.}

\edit{After an average of $1/h$-day incubation period, the exposed $E_H$ develops sufficient numbers of the transmissible form of the parasite ($\ge 10/\mu l$) in the bloodstream and becomes infectious to mosquitoes. The infectious population either develops severe disease (such as fever symptoms) $D_H$ with probability $\rho$ or remains asymptomatic $A_H$ with probability $1-\rho$. We assume that the asymptomatic individuals $A_H$ are less transmissible to mosquitoes than those who are symptomatic $D_H$, that is $\beta_A < \beta_D$. }

Recovery from severe disease $D_H$ occurs at rate $r_D$, where a portion, $\phi$, recovers to be susceptible and the rest becomes asymptomatic. Recovery from asymptomatic disease to the susceptible stage occurs at rate $r_A$, which results in clearance of the parasites. In the asymptomatic stage, $A_H$, re-exposure can cause ``superinfection'', with probability $\psi$, which results in severe disease, $D_H$. \edit{Parameters were chosen with the focus of the model being \textit{P. falciparum} malaria in Sub-Saharan Africa (\cref{tab:parameters}).} 

Transitions between disease states depend on the anti-disease immunity within the population. In particular, we assume that the transition probabilities discussed above, $\rho, \phi, \psi$, depend on the average anti-disease immunity level per person at age $\alpha$ and time $t$, i.e., $\widetilde{C}_H(\alpha,t) := C_H(\alpha,t)/P_H(\alpha,t)$. This immunity level determines the probability through sigmoid-shaped linking functions, which are described and calibrated in \cref{sec:cali}. We consider immunity on a per-person basis rather than at a population level so that immunity is not dependent on demographic effects, such as population growth.

Humans are born susceptible at rate $g_H$ and die naturally at rate $\mu_H$. People in the severe disease state, $D_H$, may suffer from the disease-induced mortality at rate $\mu_D$. For simplicity, we set $\mu_D=0$ throughout, but evaluating the impact of disease-induced mortality on our model would be an interesting future direction. 

Anti-parasitic immunity cannot be generated by infection, so we model this type of immunity solely as arising from vaccination such as from the recently approved RTS,S vaccine \cite{penny2015timecourse}. Via vaccination, the susceptible human population which received all three doses of RTS,S transitions to a vaccinated class, $V_H$, at an age-dependent vaccination rate, $\nu_p(\alpha)$, and with \edit{an initial vaccine efficacy, $\eta(\alpha)$}. The vaccinated population loses immunity, returning to the susceptible state, at rate $w$.

The following system of PDEs describes the infection dynamics in humans:
\begin{equation}
\begin{aligned}\label{eq.basic_model_human}
\partial_t S_H + \partial_\alpha S_H &= \phi(\widetilde{C}_H) r_D D_H  + r_A A_H -\Lambda_H(t)S_H\\
& \qquad -\eta (\alpha)\,\nu_p(\alpha) S_H+w V_H -\mu_H(\alpha)S_H,  \\
\partial_t E_H + \partial_\alpha E_H &= \Lambda_H(t)S_H - h E_H - \mu_H(\alpha)E_H, \\
\partial_t A_H + \partial_\alpha A_H &= (1-\rho(\widetilde{C}_H)) h E_H - \psi(\widetilde{C}_H)  \Lambda_H(t) A_H  \\ &\qquad+ (1- \phi(\widetilde{C}_H)) r_D D_H- r_A A_H - \mu_H(\alpha) A_H, \\
\partial_t D_H + \partial_\alpha D_H  &=  \rho(\widetilde{C}_H) h E_H + \psi(\widetilde{C}_H)  \Lambda_H(t) A_H \\ 
&\qquad- r_D D_H -( \mu_H(\alpha) + \mu_D(\alpha)) D_H , \\
\partial_t V_H + \partial_\alpha V_H &= \eta(\alpha)\,\nu_p(\alpha) S_H-w V_H -\mu_H(\alpha) V_H,
\end{aligned}
\end{equation}
with the boundary conditions

\begin{equation}\label{eq.BC_vh}
S_H(0,t)\! =\! \edit{\dint_0^A} \!g_H(\alpha) P_H(\alpha, t)\, d\alpha,  E_H(0,t) = A_H(0,t)  = D_H(0,t)  = V_H(0,t)  = 0,
\end{equation}
and initial conditions
\begin{align*}
S_H(\alpha,0) &= S_{H,0}(\alpha), \quad E_H(\alpha,0) = E_{H,0}(\alpha),\quad  A_H(\alpha,0)  = A_{H,0}(\alpha),\\
D_H(\alpha,0) &= D_{H,0}(\alpha), \quad V_H(\alpha,0)  = V_{H,0}(\alpha), \quad \edit{\alpha\in[0,A).}
\end{align*}
The force of infection in \cref{eq.basic_model_human} is given by
\begin{align}
\Lambda_H(t) &= b_H\big(N_M(t),N_H(t)\big)\, \beta_M\, \dfrac{I_M(t)}{N_M(t)}, \label{FOI_H_revised}
\end{align}
and the biting rate $b_H$ is defined in \cref{eq.biting_rates}. The boundary condition \cref{eq.BC_vh} assumes that all newborns are susceptible and do not receive vaccination at age zero for biological realism, \edit{thus $\nu_p(0) = 0$}.

\subsection{Mosquito Equations}\label{sec:model_mosquito}
We model the infection dynamics in mosquitoes using an ODE system, where the mosquito population is divided into three compartments: susceptible, $S_M$; exposed, $E_M$; and infectious, $I_M$. We do not include a recovered state as mosquitoes do not recover within their lifespan. The force of infection acting on mosquitoes, $\Lambda_M$, is given by \cref{FOI_M_revised} and depends on the number of bites a mosquito takes per time unit, $b_M$\edit{;} the infectivity of human infectious stages \edit{per bite}, $\beta_D$ and $\beta_A$\edit{;} and the infection level in the human population. Once infected, mosquitoes transition from exposed to infectious after $1/\sigma$ days on average. Mosquitoes are recruited at rate $g_M$ and die at rate $\mu_M$, regardless of infectious status.

We thus have the following susceptible-exposed-infected mosquito dynamics:
\begin{equation}\label{eq.basic_model_mosquito}
\begin{aligned}
\frac{dS_M}{dt}&= - \Lambda_M(t) S_M +g_M -\mu_M S_M,\\
\frac{dE_M}{dt}&= \Lambda_M(t) S_M -\sigma E_M-\mu_M E_M,\\
\frac{dI_M}{dt}&= \sigma E_M - \mu_M I_M,\quad \mbox{where}
\end{aligned}
\end{equation}
\vspace*{-12pt}
\begin{equation}\label{FOI_M_revised}
\Lambda_M(t) = b_M\big(N_M(t),N_H(t)\big) \dfrac{1}{N_H(t)} \edit{\int_0^A} \bigg( \beta_D D_H(\alpha,t)+ \beta_A A_H(\alpha,t) \bigg)\,d\alpha.
\end{equation}
The mosquito dynamics operate on a shorter time-scale relative to the disease and human demographic dynamics, and our analysis focuses on equilibrium solutions. Thus, we assume henceforth that the system \cref{eq.basic_model_mosquito} is in quasi-static equilibrium. This means that the total mosquito population $N_M(t)$ is constant and given by
\[
N_M(t) := S_M(t) + E_M(t) + I_M(t) = \frac{g_M}{\mu_M},
\]
and the number of infected mosquitoes at the quasi-static equilibrium is
\begin{equation}
I^{\star}_M(t)  = \frac{g_M}{\mu_M}\cdot \frac{\sigma}{\sigma + \mu_M}\cdot \frac{\Lambda_M(t) }{\Lambda_M(t)+\mu_M} = N_M\cdot \frac{\sigma}{\sigma + \mu_M}\cdot \frac{\Lambda_M(t) }{\Lambda_M(t)+\mu_M}.\label{eq.I_bar}   
\end{equation}
The quasi-static approximation replaces $I_M$ in \cref{FOI_H_revised} by $I_M^\star$, i.e.,
\begin{equation*}
\Lambda_H^\star(t) = b_H\big(N_M(t),N_H(t)\big)\, \beta_M\, \dfrac{I_M^\star(t)}{N_M(t)}.
\end{equation*}

To model human-mosquito contacts, we assume that the total number of bites per unit time is given by the ``compromise\edit{d}'' biting rate, 
\[
b(N_M,N_H) = b_m\,N_M\,b_h\,N_H/(b_m\,N_M + b_h\,N_H),
\]
where $b_m$ and $b_h$ are the number of bites a mosquito desires given sufficient human population and the number of bites a human can tolerate, respectively~\cite{chitnis2006bifurcation}. Thus, the compromised bites per mosquito and bites per human are given by
\begin{equation}\label{eq.biting_rates}
b_M(N_M,N_H) = \frac{b_m\,b_h\,N_H}{b_m\,N_M + b_h\,N_H} \ \text{ and } \ b_H(N_M,N_H) = \frac{b_m\,b_h\,N_M}{b_m\,N_M + b_h\,N_H},
\end{equation}
respectively. This choice allows us to consider both the commonly studied ``big city'' ($N_H \gg N_M$) and ``small village'' ($N_M \gg N_H$) scenarios in a unified framework.

\subsection{Immunity Equations}\label{sec:model_immune}
Natural immunity to malaria is acquired through repeated exposure \cite{doolan2009acquired,gupta1999immunity} so we track the immunity level within the human population and study how it subsequently affects disease transmission. There are two main types of immunity to malaria: anti-disease immunity, which reduces the probability of clinical disease, and anti-parasite immunity, which is responsible for the clearance of parasite. Anti-disease immunity affects the branching probabilities in the disease progression among humans, i.e., $\rho$, $\phi$ and $\psi$ (see \cref{fig:flowchart}). We use constant rates for parameters that are related to anti-parasite immunity, such as $r_A$, $r_D$, and $w$.

Anti-disease immunity is inherited at birth via maternal antibodies, can be developed through exposure to infected mosquitoes, and can be boosted via vaccination of specific antigen targets. Let $C_m (\alpha,t)$ denote the pooled maternal-derived immunity for all people aged $\alpha$ at time $t$, $C_e (\alpha,t)$ denote the pooled exposure-acquired immunity, and $C_{\nu} (\alpha,t)$ denote the pooled vaccine-derived immunity. The total anti-disease immunity is $C_H = c_1 C_e + c_2 C_{m} +c_3 C_{\nu}$, where $c_1$, $c_2$, and $c_3$ are scaling parameters.  

Exposure-acquired immunity is boosted from exposure to infectious mosquito bites, which is modeled by a function of the force of infection $\Lambda_H$. We assume that vaccination-protected population in $V_H$ do not contribute to the boosting of exposure-acquired immunity. The vaccine-derived immunity $C_\nu$ can be boosted through vaccination at the rate $\nu_b(\alpha,t)$; we assume only those from the susceptible population can be vaccinated. We incorporate scaling parameters $c_S, \ c_E, \ c_A, \ c_D$, and $c_{\nu}$ to model different boosting efficacies.

The immunity boosting rate increases with the exposure level, but there is a refractory period after each exposure in which immunity cannot be boosted. Following \cite{griffin2010reducing}, we employ a saturation function allowing a maximum amount of boosting per time unit, which we apply to the exposure rate $\Lambda_H$ in \cref{eq:immune_cac}. In particular, we chose
\begin{equation*}
f(x) = \frac{x}{\gamma\, x + 1}, \quad \gamma \geq  0, \quad x \geq 0.
\end{equation*}
The maternal-derived, exposure-acquired, and vaccine-derived immunity wane in time with half-life periods $d_m$, $d_e$, and $d_\nu$, respectively. As pooled quantities, immunity may also be lost due to natural or disease-induced deaths (only impact $D_H/P_H$ fraction of the age-$\alpha$ people). The age-structured PDEs for the immunity dynamics are thus given by
\begin{subequations}\label{eq.basic_model_immunity}
\begin{align}
\partial_t C_e + \partial_\alpha C_e &= f(\Lambda_H)\left( c_S S_H + c_E E_H + c_A A_H + c_D D_H\right) \label{eq:immune_cac}\\ 
&\qquad- \left(\frac{1}{d_e}+\mu_H(\alpha) + \mu_D(\alpha)\frac{D_H}{P_H} \right)C_e, \nonumber\\
\partial_t C_m + \partial_\alpha C_m &= -\left(\frac{1}{d_m} + \mu_H(\alpha) + \mu_D(\alpha)\frac{D_H}{P_H}\right)C_m,\\
\partial_t C_{\nu} + \partial_\alpha C_{\nu} &= c_{\nu} \,\nu_b(\alpha) S_H -\left(\frac{1}{d_{\nu}}+\mu_H(\alpha) + \mu_D(\alpha)\frac{D_H}{P_H} \right)C_{\nu}, 
\end{align}  
\end{subequations}
with boundary conditions,
\begin{align}
C_m(0,t) &=  m_0 \edit{\dint_{0}^{A}}g_H(\alpha) \big(c_1 C_e(\alpha,t)+c_3 C_{\nu}(\alpha,t)\big)\,d\alpha,\label{eq:immune_BC} \vspace{0.5ex}\\
C_e(0,t) &= 0, ~~ C_{\nu}(0,t) = \edit{c_{\nu} \,\nu_b(0)\,S_H(0,t)},\nonumber
\end{align}
and initial conditions, 
$C_e(\alpha,0) = C_{e,0}(\alpha)$, $C_m(\alpha,0) = C_{m,0}(\alpha)$, $C_{\nu}(\alpha,0) = C_{\nu,0}(\alpha)$. In the boundary condition for maternal immunity \cref{eq:immune_BC}, newborns inherit maternal immunity with efficacy coefficient $m_0$. We also omit the contribution of maternal immunity itself by implicitly setting $c_2=0$ in the integral. This is a simplifying assumption which keeps the expression for $C_m(0,t)$ explicit. In practice, as maternal immunity decays quickly, it is effectively zero by the time child-bearing age is reached.
\begin{table}[ht]
\caption{Description of variables used in the equations \cref{eq.basic_model_human,eq.basic_model_mosquito,eq.basic_model_immunity}. }
\label{tab:variables}
\centering
\begin{tabular}{cl}
\hline
Notation & Description \\
\hline
$S_H (\alpha, t)$ & Age density of susceptible humans at time $t$ \\
$E_H (\alpha, t)$ & Age density of exposed humans at time $t$\\
$A_H (\alpha, t)$ & Age density of asymptomatically infected humans at time $t$ \\
$D_H (\alpha, t)$ & Age density of humans with severe disease at time $t$ \\
$V_H (\alpha, t)$ & Age density of humans fully protected by vaccination\\
$P_H(\alpha, t)$ & $=S_H+E_H+A_H+D_H+V_H$, age density of humans at time $t$\\
$N_H (t)$ & $= \edit{\int_0^A} P_H(\alpha,t) d\alpha$, total humans at time $t$ \\
\hline
$S_M (t)$ & Number of susceptible mosquitoes at time $t$\\
$E_M (t)$ & Number of exposed mosquitoes at time $t$\\
$I_M (t)$ & Number of infectious mosquitoes at time $t$\\ 
$N_M (t)$ & $=S_M +E_M+I_M$, total mosquitoes at time $t$ \\
\hline
$C_e (\alpha, t)$ & Pooled exposure-acquired immunity for all people at age $\alpha$ and time $t$ \\
$C_m (\alpha, t)$ & Pooled maternal-derived immunity for all people at age $\alpha$ and time $t$ \\
$C_\nu (\alpha, t)$ & Pooled vaccine-derived immunity for all people at age $\alpha$ and time $t$ \\
$C_{H} (\alpha, t)$ & $=c_1 C_e + c_2 C_{m} +c_3 C_\nu$, total pooled anti-disease immunity  \\
$\widetilde{C}_{H} (\alpha, t)$ & $=C_H/P_H$, per-person anti-disease immunity  \\
\hline
$\Lambda_H (t)$ & Average force of infection on humans \\
$\Lambda_M (t)$ & Average force of infection on mosquitoes \\    
$f(\Lambda_H)$ & Average boosting rate of exposure-acquired immunity\\
\hline  
\end{tabular}
\end{table}
\begin{table}[ht]
\caption{Parameters and their values \edit{for Plasmodium falciparum malaria in sub-Saharan Africa} 
(see \cref{sec:cali} for calibrated parameters). 
Note in the \edit{column} for references the - - indicates the value was assumed.}
\label{tab:parameters}
\centering\small
\begin{tabular}{p{1.1cm}@{}p{7.1cm}ll@{}l@{}}
\hline
 & Description & Unit & Value & Ref.\\
\hline
$g_H(\alpha)$ & Per capita birth rate of humans&  day$^{-1}$ & dist. & \cite{kenyanationalbureauofstatistics2015kenya}\\
$\mu_H(\alpha)$ & Per capita natural mortality rate of humans & day$^{-1}$  & dist. & calibration\\
$\mu_D(\alpha)$ & Per capita disease induced mortality rate & day$^{-1}$  & 0 & - -  \\
$1/h$ & Mean incubation period in humans & day & 15
&  \cite{paaijmans2013temperaturedependent} \\
$\phi(C_H)$ & Probability of progression from $D_H$ to $S_H$ & - & - & Eq.~\cref{eq.link}\\
$\rho(C_H)$ & Probability of progression from $E_H$ to $D_H$ & - & - & Eq.~\cref{eq.link}\\
$\psi(C_H)$ & Probability of progression from $A_H$ to $D_H$ & - & - & Eq.~\cref{eq.link}\\ 
$r_A$ & Recovery rate from $A_H$ to $S_H$ & day$^{-1}$ & 1/360 & \cite{filipe2007determination}  \\
$r_D$ & Recovery rate of $D_H$ to $S_H$& day$^{-1}$ & 1/180 & \cite{filipe2007determination} \\
$g_M$ & Per capita recruitment rate of mosquitoes & day$^{-1}$ & 0.5 & - -\\
$\mu_M$ & Per capita natural mortality rate of mosquitoes & day$^{-1}$ & 1/10 & \cite{garrett-jones1964assessment}\\
$1/\sigma$ & Mean incubation period in mosquitoes & day & 15
& \cite{filipe2007determination,paaijmans2013temperaturedependent}\\
$b_h$ & Number of mosquito bites a human tolerates & day$^{-1}$ & 5 & \cite{chitnis2006bifurcation} \\
$b_m$ & Number of bites a mosquito desires & day$^{-1}$ & 0.6 & \cite{chitnis2006bifurcation}\\    
$b_H(\cdot,\cdot)$ & Number of mosquito bites per person  & day$^{-1}$ & - & Eq.~\cref{eq.biting_rates}\\
$b_M(\cdot,\cdot)$ & Number of bites per mosquito & day$^{-1}$ & - & Eq.~\cref{eq.biting_rates} \\
$\beta_M$ & \edit{Per bite infectivity} of infectious mosquitoes $I_M$ & - & 0.25 & \cite{filipe2007determination}\\
$\beta_D$ & \edit{Per bite infectivity} of humans with severe disease & - & 0.35 & \cite{filipe2007determination}\\
$\beta_A$ & \edit{Per bite infectivity} of asymptomatic humans & - & 0.03 & \cite{filipe2007determination}\\
$d_m$ & Average length of maternal immunity & year & 0.25 & \cite{filipe2007determination,snow1998risk} \\
$d_e$ & Average length of exposure-acquired immunity & year & 5 & \cite{filipe2007determination,yman2019antibody} \\
\hline
\edit{$\nu_b (\alpha)$} & Vaccination (immunity boosting) rate & year$^{-1}$ & dist. & - -\\
\edit{$d_{\nu}$}& Average length of vaccine-boosted immunity ($C_\nu$) & year & 5 & - - \\
\hline
\edit{$\nu_p (\alpha)$}\!\! & Vaccination rate  & year$^{-1}$ & dist. & - - \\
\edit{$\eta(\alpha)$} & Vaccine efficacy against infection & - & 0.73 & \cite{penny2015timecourse}\\
$w$ & Waning rate of infection-protection immunity & year$^{-1}$ & 1/0.66 & \cite{penny2015timecourse} \\
\hline
& Immunity acquisition coefficients &&\\
\hline
$m_0$ & Fraction of maternal immunity conferred & - & 1 & - -\\
$c_1$ & Weight for exposure-acquired immunity & - & 1 & - -\\
$c_2$ & Weight for maternal immunity & - & 1 & - -\\
$c_3$ & Weight for vaccine-derived immunity & - & 1 & - -\\
$c_S$ & Weight for boosting at $S_H$  & - & 0.75 & - -\\
$c_E$ & Weight for boosting at $E_H$  & - & 0.1 & - -\\
$c_A$ & Weight for boosting at $A_H$  & - & 0.1 & - -\\
$c_D$ & Weight for boosting at $D_H$  & - & 0.05 & - -\\
$c_{\nu}$ & \edit{Relative strength of} boosting from vaccination  & - & 0.75 &  - -\\
\hline
\end{tabular}
\end{table}

\section{Well-posedness of the Model}
The human-mosquito-immunity coupled system given by \cref{eq.basic_model_human,eq.basic_model_mosquito,eq.basic_model_immunity} is well-posed and has a unique nonnegative solution under biologically reasonable conditions on the coefficients, parameters, and the initial and boundary conditions. We assume for simplicity that the human population size, $N_H$, is constant. To achieve constant population size, suppose
\begin{equation}\label{eq.P_H_choice}
\edit{P_H^*(\alpha) = \mu_H^* \, N_H\, e^{-M(\alpha)}}, \quad \mu_H^* := \left(\edit{\int_0^A} e^{-M(\alpha)} \,d \alpha\right)^{-1}, \quad N_H >0,
\end{equation}

\noindent where $\mu_H^*$ is the so-called crude death rate and $M(\alpha) := \int_0^\alpha \mu_H(\sigma)\,d\sigma$. We further assume that the mosquito population is fixed at its equilibrium level, i.e., $N_M=g_M/\mu_M$. \edit{We impose the following conditions to guarantee well-posedness of the model:}
\begin{enumerate}
\item[(H1)] $S_M(0)$, $E_M(0)$ and $I_M(0)$ are nonnegative, 
\item[(H2)] $S_{H,0}$, $E_{H,0}$, $A_{H,0}$, $D_{H,0}$, $V_{H,0}$, $C_{m,0}$, $C_{e,0}$, \edit{$C_{\nu,0}$ $\in L^1((0,A);\mathbb{R}_+)$},
\item[(H3)] \edit{The parameters $d_m$, $d_e$, $d_\nu$, $g_M$, $\mu_M$, $h$, $r_A$, $r_D$ and $w$ are positive, and all other parameters are non-negative,}
\item[(H4)] \edit{$\mu_H \in L^1_{loc}((0,A);\mathbb{R}_+)$ with $\edit{\int_0^A} \mu_H(\alpha)\,d\alpha=\infty$},
\item[(H5)] \edit{$f \in \mbox{Lip}(\mathbb{R}_+;\mathbb{R}_+)$, $g_H$, $\nu_p$, $\nu_b$, $\eta \in L^\infty((0,A); \mathbb{R}_+)$,  $\mbox{essinf}_{\alpha\in[0,A)}\eta(\alpha)\nu_p(\alpha)>0$}.
\item[(H6)] $\phi,\,\psi,\,\rho \in \mbox{Lip}(\mathbb{R}_+;[0,1])$.
\item[(H7)] There is no disease induced mortality, i.e., $\mu_D \equiv 0$.
\end{enumerate}

\edit{It is usually straightforward to establish appropriate existence and uniqueness result for a mixed ODE-PDE age-structured system and numerous results in the literature do so via semigroup methods~\cite{inaba1990threshold,webb1985theory}.
However, due to the nature of the boundary conditions in our problem, we must proceed via the results of Thieme~\cite{thieme1990semiflows} instead (see also \cite{cai2013epidemic,martcheva2003progression} for similar applications to age-structured models). Full details regarding the well-posedness of the model are supplied in \cref{sec_appendix_wellposed}.}

\section{Stability of the Disease Free Equilibrium (DFE) and \texorpdfstring{$\mathcal{R}_0$\ }\ Calculation}\label{sec_R0}
\subsection{\edit{Normalized System}}
\edit{
In order to simplify the forthcoming stability calculations, we rewrite the system using population proportions, which are given by
$\widetilde{S}_H(\alpha,t)\!=\!S_H(\alpha,t)/P_H(\alpha,t)$, $\widetilde{E}_H(\alpha,t) = E_H(\alpha,t)/P_H(\alpha,t)$, and so on. The proportions thus obey the following evolution equations:
\begin{equation}\label{eq.immunity_proportions}
\begin{array}{l}
\partial_t\widetilde{S}_H+\partial_\alpha \widetilde{S}_H = -\Lambda_H^\star(t)\widetilde{S}_H \!+\! \phi(\widetilde{C}_H ) r_D\, \widetilde{D}_H \!+\! r_A \widetilde{A}_H\!-\! \edit{\eta(\alpha)\nu_p(\alpha)  \widetilde{S}_H}\!+\!w \widetilde{V}_H,\\
\partial_t\widetilde{E}_H+\partial_\alpha \widetilde{E}_H = \Lambda_H^\star(t)\widetilde{S}_H - h \widetilde{E}_H,\\
\partial_t\widetilde{A}_H+\partial_\alpha \widetilde{A}_H= (1-\rho(\widetilde{C}_H )) h \widetilde{E}_H -( \psi(\widetilde{C}_H) \Lambda_H^\star(t) + r_A) \widetilde{A}_H \\ 
\hspace{0.3\textwidth} + (1- \phi(\widetilde{C}_H )) r_D \widetilde{D}_H,  \\
\partial_t\widetilde{D}_H+\partial_\alpha \widetilde{D}_H  =  \rho(\widetilde{C}_H ) h \widetilde{E}_H + \psi(\widetilde{C}_H) \Lambda_H^\star(t) \widetilde{A}_H - r_D \widetilde{D}_H,  \\
\partial_t\widetilde{V}_H+\partial_\alpha \widetilde{V}_H = \eta(\alpha)\,\nu_p(\alpha) \widetilde{S}_H-w \widetilde{V}_H,\\
\widetilde{S}_H(0,t) = 1,  \quad \widetilde{E}_H(0,t)  = 0, \quad \widetilde{A}_H(0,t)  = 0,\quad \widetilde{D}_H(0,t)  = 0, \quad \widetilde{V}_H(0,t)  = 0.
\end{array}
\end{equation}
The system \cref{eq.immunity_proportions} has the same force of infection $\Lambda_H(t)$ as defined in \cref{FOI_H_revised}, and under the quasi-static approximation \cref{eq.I_bar}, it may be written as
\[
\Lambda^\star_H(t) = b_H(N_M,N_H)\cdot\frac{\beta_M\,\sigma}{\sigma+\mu_M}\cdot \frac{\Lambda_M(t)}{\Lambda_M(t)+\mu_M},\quad \mbox{where}\] 
\vspace*{-16pt}
\[
 \Lambda_M(t) = b_M(N_M,N_H)\,\mu_H^* \edit{\int_0^A} e^{-M(\alpha)}\left(\beta_D \widetilde{D}_H(\alpha,t) + \beta_A \widetilde{A}_H(\alpha,t) \right)\,d\alpha.\label{eq.FOI_M_props} 
\]
The exposure, maternal and vaccine-derived immunity per person evolve according to
\begin{equation}\label{eq.immunity_proportions2}
\begin{array}{cl}
\partial_t \widetilde{C}_e + \partial_\alpha \widetilde{C}_e &= f(\Lambda^\star_H)\left\{ c_S \widetilde{S}_H + c_E \widetilde{E}_H + c_A \widetilde{A}_H + c_D \widetilde{D}_H \right\}- \dfrac{\widetilde{C}_e}{d_e},\\
\partial_t \widetilde{C}_m + \partial_\alpha \widetilde{C}_m &= - \dfrac{\widetilde{C}_m}{d_m},\\
\partial_t \widetilde{C}_{\nu} + \partial_\alpha \widetilde{C}_{\nu} &=  c_{\nu} \,\nu_b(\alpha) \widetilde{S}_H - \dfrac{\widetilde{C}_{\nu}}{d_{\nu}},
\end{array}
\end{equation}
with initial and boundary conditions, respectively, given by
\begin{align*}
\widetilde{C}_e(0,t) &=0 , ~\widetilde{C}_m(0,t) =  \frac{m_0}{P_H(0,t)} \edit{\int_{0}^{A}}g_H P_H \big(c_1\widetilde{C}_e +c_3\widetilde{C}_{\nu}\big)\,d\alpha, ~\widetilde{C}_{\nu}(0,t) =c_{\nu} \nu_b(0); \\ 
\widetilde{C}_e(\alpha,0) &= \widetilde{C}_{e,0}(\alpha),~~ \widetilde{C}_m(\alpha,0) = \widetilde{C}_{m,0}(\alpha),~~\widetilde{C}_\nu(\alpha,0) = \widetilde{C}_{\nu,0}(\alpha).
\end{align*}
Total immunity per person is given by
$
\widetilde{C}_{H}(\alpha,t) = c_1 \widetilde{C}_e(\alpha,t) + c_2 \widetilde{C}_m(\alpha,t) + c_3 \widetilde{C}_{\nu}(\alpha,t).
$ 
}

\subsection{Stability of the Disease Free Equilibrium}
In addition to regularity conditions (H1)-(H7), \edit{for the stability analysis, we further assume that $\rho$, $\psi$, $\phi$ are smooth, i.e., $\rho$, $\psi$, $\phi \in C^\infty\left(\mathbb{R}_+;[0,1]\right)$.}

Since disease-induced mortality $\mu_D$ is absent \edit{by assumption}, summing the human compartment equations \cref{eq.basic_model_human} shows that the number of age $\alpha$ humans at time $t$, $P_H(\alpha,t)$, evolves according to
\begin{equation}\label{eq.P_H_PDE}
\partial_t P_H + \partial_{\alpha} P_H = -\mu_H(\alpha)P_H, \quad P_H(0,t) = \edit{\int_0^A} g_H(\alpha) P_H(\alpha, t)\,d\alpha.
\end{equation}
Equation \cref{eq.P_H_PDE} can be solved explicitly using the method of characteristics~\cite{m1925applications}, and its behaviour as $t \to \infty$ is given by
\begin{equation}\label{eq.stable_age}
P_H(\alpha,t) \approx K e^{-q\alpha -M(\alpha)}\, e^{qt},\quad K = N_H\left(\edit{\dint_0^A} e^{-M(a)-q a} \,da \right)^{-1}\in \mathbb{R}_+,
\end{equation}
where $N_H\in\mathbb{R}_+$ is the initial population size, and the constant $q$ solves
\begin{equation}\label{eq.q_zero}
\edit{\int_0^A} g_H(\alpha)e^{-q\alpha-M(\alpha)}\,d\alpha = 1,
\end{equation}
(see, for example, \cite{allen2008mathematical}). The term $K e^{-q\alpha-M(\alpha)}$ in \cref{eq.stable_age} is the \emph{stable-age distribution} which is the asymptotically stable demographic structure of the population. If $q<0$, the population will eventually die out; if $q>0$, the population grows without bound. Hence we assume that $q=0$, so that the human population has both a constant size and a constant demographic structure. \edit{If we start from the stable age distribution with $q=0$, then \cref{eq.P_H_choice} holds, i.e., $P_H^*(\alpha) = \mu_H^* \, N_H\, e^{-M(\alpha)}$. }

The DFE of the system \cref{eq.immunity_proportions,eq.immunity_proportions2} has the form 
\begin{align*}
&\left\{\left(\theta(\alpha),0,0,0,1-\theta(\alpha),\widetilde{C}_e^*(\alpha),\widetilde{C}^*_m(\alpha),\widetilde{C}_{\nu}^*(\alpha)\right)\,:\, \alpha \geq 0\right\},\quad \mbox{where}\\
\theta(\alpha) &=\edit{e^{-\int_0^\alpha \pi(a)\,da}}\left(1+\int_0^\alpha w \,e^{\int_0^a \pi(z)\,dz}\,da\right),\quad \pi(\alpha)=w+\edit{\eta(\alpha)}\,\nu_p(\alpha).
\end{align*}
For any level of $\widetilde{C}_{H}$, the disease free solution $(\theta(\alpha),0,0,0,1-\theta(\alpha))$ is an equilibrium of the human disease compartments subsystem \cref{eq.immunity_proportions}. It remains to identify the steady state immunity distributions $\widetilde{C}_e^*$, $\widetilde{C}^*_m$, and $\widetilde{C}_{\nu}^*$ at the DFE. At the DFE, the steady state for vaccine-derived immunity obeys \edit{$
\frac{d}{d\alpha} \widetilde{C}_{\nu}^*(\alpha) = - \widetilde{C}_{\nu}^*/d_{\nu} + c_{\nu} \nu_b(\alpha)\theta(\alpha)
$} with initial condition $\widetilde{C}_{\nu}^*(0) = c_{\nu} \nu_b(0)$. Hence
\[
\widetilde{C}_{\nu}^*(\alpha) = c_{\nu} \, e^{-\alpha/d_{\nu}}\left(\nu_b(0) + \edit{\int_0^\alpha e^{a/d_{\nu}} \nu_b(a)\theta(a) \,da} \right).
\]
Exposure-acquired immunity obeys 
$
\frac{d}{d\alpha} \widetilde{C}_e^*(\alpha) = - \widetilde{C}_e^*/d_e
$ with $\widetilde{C}_e^*(0) = 0$. Hence $\widetilde{C}_e^*(\alpha) \equiv 0$.
Similarly, the steady state for maternal immunity obeys
\[
\frac{d}{d\alpha} \widetilde{C}_{m}^*(\alpha) = - \frac{\widetilde{C}_{m}^*}{d_{m}}, \quad \widetilde{C}_m^*(0) = m_0 \edit{\int_0^A} g_H(\alpha)e^{-M(\alpha)}\big(c_3\widetilde{C}_{\nu}^*(\alpha)\big)\,d\alpha.
\]
Thus, the maternal immunity steady state has the form $
\widetilde{C}_{m}^*(\alpha) = \widetilde{C}_m^*(0) e^{-\alpha/d_m}.$ 
Finally, the steady state distribution for total immunity per person is given by
\[
\widetilde{C}^*_{H}(\alpha) = c_2 \,\widetilde{C}_m^*(0) e^{-\alpha/d_m}+ c_3 c_{\nu} e^{-\alpha/d_{\nu}}\left(\nu_b(0) + \edit{\int_0^\alpha e^{a/d_{\nu}} \nu_b(a)\theta(a) \,da} \right).
\]
To determine the stability of the DFE, consider a perturbation of the form
\begin{align*}
\widetilde{S}_H(\alpha,t) &= \theta(\alpha) + \varepsilon_S(\alpha,t), \quad \widetilde{E}_H(\alpha,t) = \varepsilon_E(\alpha,t), \quad \widetilde{A}_H(\alpha,t) = \varepsilon_A(\alpha,t),\\
\widetilde{D}_H(\alpha,t) &= \varepsilon_D(\alpha,t),\quad 
\widetilde{V}_H(\alpha,t) = 1-\theta(\alpha)+\varepsilon_V(\alpha,t),\\ 
\widetilde{C}_e(\alpha,t) & = \varepsilon_e(\alpha,t), ~~  \widetilde{C}_{m}(\alpha,t) = \widetilde{C}^*_{m}(\alpha) + \varepsilon_{m}(\alpha,t), ~~\widetilde{C}_{\nu}(\alpha,t)  = \widetilde{C}^*_{\nu}(\alpha) +
\varepsilon_{\nu}(\alpha,t).
\end{align*}
For notational convenience, let $\widetilde{C}_{H}(\alpha,t) = \widetilde{C}_{H}^*(\alpha) + \varepsilon_{H}(\alpha,t)$ and $\varepsilon_{H}(\alpha,t) = c_1 \varepsilon_e(\alpha,t) + c_2 \varepsilon_{m}(\alpha,t)+c_3 \varepsilon_{\nu}(\alpha,t)$. First, linearizing the force of infection yields
\[
\Lambda^\star_H(t) \approx C^\star \edit{\int_0^A} e^{-M(\alpha)}\left( \beta_D \varepsilon_D(\alpha,t)+\beta_A \varepsilon_A(\alpha,t)\right)d\alpha,
\]
where $C^\star = \mu_H^*\, b_m^2\,b_h^2\,N_M\,N_H\,\beta_M\,\sigma/(b_m\,N_M + b_h\,N_H)^2(\sigma+\mu_M)\mu_M
$. Next linearize the system of PDEs \cref{eq.immunity_proportions}. We can first linearize the terms not involving immunity to reduce the human subsystem to
\begin{align*}
{\partial_t \varepsilon_S}+{\partial_\alpha \varepsilon_S} &= - \theta(\alpha)\Lambda^\star_H(t) + \phi(\widetilde{C}_{H}) r_D \varepsilon_D + r_A \varepsilon_A - \eta(\alpha) \nu_p(\alpha)\varepsilon_S+w \varepsilon_V,\\
{\partial_t \varepsilon_E}+{\partial_\alpha \varepsilon_E} &= \theta(\alpha)\Lambda_H^\star(t) - h \varepsilon_E,\\
{\partial_t \varepsilon_A}+{\partial_\alpha \varepsilon_A} &= (1-\rho(\widetilde{C}_{H})) h \varepsilon_E + (1- \phi(\widetilde{C}_{H})) r_D \varepsilon_D  - r_A \varepsilon_A,\\
{\partial_t \varepsilon_D}+{\partial_\alpha \varepsilon_D}  &=  \rho(\widetilde{C}_{H}) h \varepsilon_E - r_D \varepsilon_D,  \\
{\partial_t \varepsilon_V}+{\partial_\alpha \varepsilon_V}  &=  \eta(\alpha) \nu_p(\alpha)\varepsilon_S-w \varepsilon_V.
\end{align*}
We linearize the immunity terms as follows. Expanding $\phi(\widetilde{C}_{H}) = \phi(\widetilde{C}^*_{H} + \varepsilon_{H})$ about the steady state $\widetilde{C}^*_{H}(\alpha)$ at a given age $\alpha$ to obtain 
$\phi(\widetilde{C}_{H}) \approx \phi(\widetilde{C}^*_{H}) + \phi'(\widetilde{C}^*_{H})\varepsilon_{H} + {\scriptstyle \mathcal{O}}(\varepsilon_{H}^2)$. Thus, to leading order in the perturbations, we obtain $\phi(\widetilde{C}_{H}) r_D \varepsilon_D \approx \phi(\widetilde{C}^*_{H}) r_D \varepsilon_D$. Treating the other immunity terms similarly yields
\begin{align*}
{\partial_t \varepsilon_S}+{\partial_\alpha \varepsilon_S} &= -\theta(\alpha) \Lambda^\star_H(t) + \phi(\widetilde{C}^*_{H}) r_D \varepsilon_D + r_A \varepsilon_A - \eta(\alpha) \nu_p(\alpha)\varepsilon_S+w \varepsilon_V, \\
{\partial_t \varepsilon_E}+{\partial_\alpha \varepsilon_E} &= \theta(\alpha) \Lambda^\star_H(t) - h \varepsilon_E,\\
{\partial_t \varepsilon_A}+{\partial_\alpha \varepsilon_A} &= (1-\rho(\widetilde{C}_{H}^*)) h \varepsilon_E + (1- \phi(\widetilde{C}^*_{H})) r_D \varepsilon_D  - r_A \varepsilon_A,\\
{\partial_t \varepsilon_D}+{\partial_\alpha \varepsilon_D}  &=  \rho(\widetilde{C}_{H}^*) h \varepsilon_E - r_D \varepsilon_D,\\
{\partial_t \varepsilon_V}+{\partial_\alpha \varepsilon_V}  &=  \eta(\alpha) \nu_p(\alpha)\varepsilon_S-w \varepsilon_V.
\end{align*}
Next make the exponential ansatz $\varepsilon_{\cdot}(\alpha,t) = e^{pt}\hat{\varepsilon}_{\cdot}(\alpha)$ for some $p \in \mathbb{C}$, to reduce the linearized system to the following set of ODEs:
\begin{subequations}
\begin{align}
{\frac{d}{d\alpha} \hat{\varepsilon}_S} &= -\theta(\alpha) \hat{\Lambda}_H  + \phi(\widetilde{C}^*_{H}(\alpha)) r_D \hat{\varepsilon}_D + r_A \hat{\varepsilon}_A  - (\eta(\alpha) \nu_p(\alpha)+p)\hat{\varepsilon}_S+w \hat{\varepsilon}_V,\\
{\frac{d}{d\alpha} \hat{\varepsilon}_E} &= \theta(\alpha) \hat{\Lambda}_H  - (h+p)\, \hat{\varepsilon}_E, \label{eq.immunity_epsilonE}\\
{\frac{d}{d\alpha} \hat{\varepsilon}_A} &= (1-\rho(\widetilde{C}_{H}^*(\alpha)))\, h\, \hat{\varepsilon}_E + (1- \phi(\widetilde{C}^*_{H}(\alpha)))\, r_D\, \hat{\varepsilon}_D  - (r_A+p) \hat{\varepsilon}_A,\label{eq.immunity_epsilonA}\\
{\frac{d}{d\alpha} \hat{\varepsilon}_D}  &=  \rho(\widetilde{C}_{H}^*(\alpha)) h \hat{\varepsilon}_E - (r_D+p) \hat{\varepsilon}_D,  \label{eq.immunity_epsilonD}\\
{\frac{d}{d\alpha} \hat{\varepsilon}_V}  &= \eta(\alpha)\nu_p(\alpha)\hat{\varepsilon}_S-(w+p) \hat{\varepsilon}_V,
\end{align}
\end{subequations}
with initial conditions
$
\hat{\varepsilon}_S(0) = \hat{\varepsilon}_E(0) = \hat{\varepsilon}_D(0) = \hat{\varepsilon}_A(0) =\hat{\varepsilon}_V(0)= 0. 
$ 
The linearized force of infection $\hat{\Lambda}_H$ is now given by
\[
\hat{\Lambda}_H := C^\star \edit{\int_0^A} e^{-M(\alpha)}\left( \beta_D \hat{\varepsilon}_D(\alpha)+\beta_A \hat{\varepsilon}_A(\alpha)\right)d\alpha.
\]
We can immediately read off from \cref{eq.immunity_epsilonE} that
\[
\hat{\varepsilon}_E(\alpha) =  \hat{\Lambda}_H \int_0^\alpha e^{-(h+p)(\alpha-a)}\theta(a)\,da  =: \hat{\Lambda}_H\, \mathcal{E}(\alpha,p),
\]
and similarly, it follows from \cref{eq.immunity_epsilonD} that
\[
\hat{\varepsilon}_D(\alpha) = \hat{\Lambda}_H\, h \int_0^\alpha e^{-(r_D+p)(\alpha-a)} \rho(\widetilde{C}_{H}^*(a)) \mathcal{E}(a,p)\,da =: \hat{\Lambda}_H \, \mathcal{D}(\alpha,p).
\]
Finally, solving \cref{eq.immunity_epsilonA} yields the following expression for $\hat{\varepsilon}_A(\alpha)$:
\begin{multline*}
\hat{\Lambda}_H  \int_0^\alpha e^{-(r_A+p)(\alpha-a)}\left(h (1-\rho(\widetilde{C}_{H}^*(a)))\,\mathcal{E}(a,p) + r_D (1- \phi(\widetilde{C}^*_{H}(a))) \,\mathcal{D}(a,p)  \right) da \\  =: \hat{\Lambda}_H\, \mathcal{A}(\alpha,p).
\end{multline*}
Plugging the expressions above for $\hat{\varepsilon}_D$ and $\hat{\varepsilon}_A$ into the force of infection $\hat{\Lambda}_H$ leads to the following characteristic equation for $p$:
\begin{equation}\label{eq.characteristic_immunity}
1 = C^\star \edit{\int_0^A} e^{-M(\alpha)}\left( \beta_D\, \mathcal{D}(\alpha,p)+\beta_A \,\mathcal{A}(\alpha,p)\right)d\alpha =: \mathcal{\zeta}(p).
\end{equation}
The stability of the DFE is determined by the sign of the root of the nonlinear equation $\zeta(p)-1 = 0$ with the largest real part, which we denote by $p^\star$. When $\mbox{Re}(p^\star)>0$, linear perturbations will result in an exponential growth from the DFE, which indicates that the DFE is unstable. Similarly, when $\mbox{Re}(p^\star)<0$, perturbations about the DFE decay exponentially, indicating that the DFE is linearly asymptotically stable. We set $p=0$ in the right-hand side of \cref{eq.characteristic_immunity} to obtain the quantity
\begin{equation}\label{eq.R0}
\mathcal{R}_0^\star := \zeta(0) = C^\star \edit{\int_0^A} e^{-M(\alpha)}\left( \beta_D \,\mathcal{D}(\alpha,0)+\beta_A\, \mathcal{A}(\alpha,0)\right)d\alpha,
\end{equation}
which provides a threshold condition for the stability of the DFE.
\begin{theorem}\label{thm:R0threshold}
The DFE is locally asymptotically stable if $\mathcal{R}_0^\star< 1$ and unstable if $\mathcal{R}_0^\star>1$.
\end{theorem}
\edit{\begin{remark}
The rigorous justification that the roots of the characteristic equation determine the local asymptotic stability of the DFE for our model can be proven by following the arguments of Martcheva and Thieme~\cite[Appendix B]{martcheva2003progression}.
\end{remark}}
\begin{proof}
Suppose $\mathcal{R}_0^\star = \zeta(0) < 1$. For $p \in \mathbb{C}$, it can be shown that $|\zeta(p)| \leq \zeta\left( \text{Re}(p) \right)$ and furthermore, for $p \in \mathbb{R}_+$, $p \mapsto \zeta(p)$ is non-increasing (see \cref{sec:appendixR0} for further details). If the real part of $p$ is positive, then
\[
|\zeta(p)| \leq \zeta\left( \text{Re}(p) \right) \leq \zeta(0) = \mathcal{R}_0^\star < 1, \quad p \in \mathbb{C},
\]
a contradiction. Therefore the characteristic equation $\zeta(p) = 1$ cannot have a solution with positive real part.

For $p \in \mathbb{R}_+$, $p \mapsto \zeta(p)$ is continuous. It can be shown that 
\[
\lim_{p \to \infty}\mathcal{D}(\alpha,p) = \lim_{p \to \infty}\mathcal{A}(\alpha,p) = 0\quad (\text{uniformly in $\alpha$})
\]
and hence that $\lim_{p\to\infty}\zeta(p) = 0$ (see \cref{sec:appendixR0}). If $\mathcal{R}_0^\star = \zeta(0) > 1$, then $\zeta(0)-1>0$ and for $p$ sufficiently large, $\zeta(p)-1<0$ so by continuity the characteristic equation has at least one positive real root. Therefore the DFE is unstable in this case.
\end{proof}

The expression given in \eqref{eq.R0} is a two-generation factor \cite{diekmann1990definition}; we thus define the \textit{basic reproduction number} as the average next generation factor,
\begin{equation}\label{eq.R02}
\mathcal{R}_0 := \sqrt{\mathcal{R}_0^\star}.    
\end{equation}
\subsection{\texorpdfstring{$\mathcal{R}_0$\ }\  Interpretation} 
We rewrite $\mathcal{R}_0$ in \cref{eq.R02} as follows,
\begin{align*} 
\mathcal{R}_0 &= \sqrt{ \left(b_M\, \beta_M\, \frac{\sigma}{\sigma+\mu_M}\cdot \frac{1}{\mu_M} \right)\!\times\!
\left(b_H  \edit{\int_0^A} \mu_H^*\, e^{-M(\alpha)}\left( \beta_D \,\mathcal{D}(\alpha,0)+\beta_A\, \mathcal{A}(\alpha,0)\right)d\alpha\right)}\\
&=: \sqrt{\mathcal{R}_{MH}\times \mathcal{R}_{HM}},
\end{align*}
where $\mathcal{R}_{MH}$ and $\mathcal{R}_{HM}$ are two reproduction numbers for the two one-way transmission routes:
One infectious human (mosquito) may generate $\mathcal{R}_{HM}$ ($\mathcal{R}_{MH}$) infected mosquitoes (humans) per generation, and the number of new infectious individuals created throughout one complete cycle (two generations) is the product of the two one-way reproduction numbers. The overall basic reproduction number per generation for one infectious individual, regardless of whether it's a mosquito or human, is the geometric mean of $\mathcal{R}_{HM}$ and $\mathcal{R}_{MH}$.

\paragraph{Mosquito-to-human transmission}
The reproduction number for\\ mosquito-to-human transmission is given by
\[
\mathcal{R}_{MH}= b_M\, \beta_M\, \frac{\sigma}{\sigma+\mu_M}\frac{1}{\mu_M}.
\]
An infected mosquito enters $I_M$ and survives to become infectious with probability $\sigma/(\sigma+\mu_M)$, and it spends on average $1/\mu_M$ being infectious. Thus the expected number of human infections generated by an infected mosquito, $\mathcal{R}_{MH}$, is the product of the number of bites a mosquito has per day, $b_M$, the probability of transmission to human per bite, $\beta_M$, and the expected infectious period of a mosquito, $\tau_M = \sigma/(\sigma+\mu_M)\mu_M$. 
\paragraph{Human-to-mosquito transmission}
The reproduction number for \\
human-to-mosquito transmission is given by
\begin{equation}\label{eq.RHM}
\mathcal{R}_{HM} = b_H  \edit{\int_0^A} \mu_H^*\, e^{-M(\alpha)}\left( \beta_D \,\mathcal{D}(\alpha,0)+\beta_A\, \mathcal{A}(\alpha,0)\right)d\alpha,\quad \mbox{where}
\end{equation}
\vspace*{-12pt}
\begin{subequations}
\begin{align}
\mathcal{E}(\alpha,0) & = \int_0^\alpha e^{-h(\alpha-a)}\theta(a)\,da, \label{eq.E} \\
\mathcal{D}(\alpha,0) & =  \int_0^\alpha e^{-r_D(\alpha-a)} \rho \,h\,\mathcal{E}(a,0)\,da, \label{eq.F}\\
\mathcal{A}(\alpha,0) & = \int_0^\alpha  e^{- r_A (\alpha-a)}\Big(h (1-\rho)\,\mathcal{E}(a,0) + r_D (1- \phi) \,\mathcal{D}(a,0)  \Big) \,da. \label{eq.G}
\end{align}
\end{subequations}
To facilitate the interpretation, let $X(\alpha)$ denote the infection status of a person at age $\alpha$, and define $T_s$ as the time that the person spends in stage $s$, which follows an exponential distribution with parameter given by the rate of transition out of $s$. 

At the DFE, the probability that a randomly chosen person of age $a$ can be infected (i.e., susceptible, not protected by vaccination) is $\theta(a)$. For an age-$\alpha$ infected person, if we assume infection occurs at age $a$, then the person must spend $\alpha-a$ days in stage $E_H$ and $T_{E_H} \sim \text{Exp}(h)$. Thus, conditioning on the age of infection, the probability that an age-$\alpha$ person is in stage $E_H$ can be written as
\[
\mathbb{P}(X(\alpha)=E_H) = \int_0^\alpha h e^{-h(\alpha-a)} \theta(a) \, da = h\,\mathcal{E}(\alpha,0).
\]
To interpret \cref{eq.F}, we consider an age-$\alpha$ person, who was in state $E_H$ at age $a$ (with probability $\mathbb{P}(X(a)=E_H)$) and immediately progressed to $D_H$ (with probability $\rho$). The person then spends $\alpha-a$ days in $D_H$ and $T_{D_H} \sim \text{Exp}(r_D)$. Thus, conditioning on the age of transition to $D_H$, the probability an age-$\alpha$ person is in $D_H$ is given by
\[
\mathbb{P}(X(\alpha)=D_H) = \int_0^\alpha  r_D \,e^{-r_D(\alpha-a)} \rho\,\mathbb{P}(X(a)=E_H) \, da=
 r_D\, \mathcal{D}(\alpha,0).
\]
We note that, in general, an infected person can also enter $D_H$ due to the superinfection ($E_H\rightarrow A_H\rightarrow D_H$), however, this situation does not happen at DFE.

On average, people spend $\tau_D := 1/r_D$ days in $D_H$ (regardless of age). Thus the expected time that an infected age-$\alpha$ person spends in $D_H$ is
\[
\mathcal{D}(\alpha,0) = \mathbb{P}(X(\alpha)=D_H)\times \tau_D.
\]

In \cref{eq.G}, an age-$\alpha$ infected person may enter $A_H$ stage via two routes, $E_H\rightarrow A_H$ or $E_H\rightarrow D_H\rightarrow A_H$. For the first scenario, suppose the person progresses from $E_H$ to $A_H$ at age $a$ and  spends $\alpha-a$ days in $A_H$, where $T_{A_H}\sim \text{Exp}(r_A)$. Thus
\begin{equation}
\mathbb{P}(X(\alpha) = A_H \, , \,E_H\rightarrow A_H) =  \int_0^\alpha \!\! r_A \,e^{-r_A(\alpha-a)}  (1-\rho)\,  \mathbb{P}(X(a)=E_H) \, da.
\label{eq.PA}
\end{equation}
Similarly, for the second route, suppose the person progresses from $D_H$ to $A_H$ at age $a$ and spends $\alpha-a$ days in $A_H$, then
\begin{equation}
\mathbb{P}( X(\alpha)\! = \!A_H \,,\, E_H\rightarrow D_H\rightarrow A_H)\! =\!\int_0^\alpha \!\! r_A \,e^{-r_A(\alpha-a)}  (1-\phi)\, \mathbb{P}(X(a)=D_H) \, da.
\label{eq.PD}
\end{equation}
Now sum \cref{eq.PA,eq.PD} to obtain the probability that an age-$\alpha$ person is in $A_H$,
\[
\mathbb{P}(X(\alpha)=A_H) = \int_0^\alpha \!\! r_A \, e^{-r_A(\alpha-a)} \Big( h \,  (1-\rho)\mathcal{E}(a,0)+ r_D\,(1-\phi) \mathcal{D}(a,0)  \Big)\, da = r_A\, \mathcal{A}(\alpha,0).
\]
If $\tau_A: = 1/r_A$ denotes the expected time people spend in stage $A_H$, then
\[
\mathcal{A}(\alpha,0) = \mathbb{P}(X(\alpha)=A_H)\times \tau_A
\]
is the expected time that an infected age-$\alpha$ person spends in $A_H$.

Finally, the expected number of infected mosquitoes created by an infected age-$\alpha$ person across all infectious states, $D_H$ and $A_H$, is given by
\[
\mathcal{R}_{HM,\alpha} := b_H\beta_D\mathcal{D}(\alpha,0) + b_H\beta_A\mathcal{A}(\alpha,0),
\]
and, by the law of total expectation, the human-to-mosquito reproduction number is
\begin{align*}
\mathcal{R}_{HM} &= \mathbb{E}\left[\mathbb{E}[\text{cases produced per infected person}\,\vert\, \text{person is aged } \alpha ]\right]\\
&= \mathbb{E}\left[ \mathcal{R}_{HM,\alpha} \right]=\edit{\int_0^A} \mathcal{R}_{HM,\alpha} \, \mu_H^*\,e^{-M(\alpha)}\, d\alpha.
\end{align*}
This recovers the reproduction number stated in \cref{eq.RHM}.

\section{Numerical Examples} 
All codes to reproduce the results of this section are available on Github \edit{(\url{https://github.com/AMSMRC1A/age_struct_malaria})}. Parameter values are shown in \cref{tab:parameters}; the time-unit for calculations is days but converted to years for plots. We assume the scaled population sizes are $N_H=1$ and $N_M = g_M/\mu_M = 5$ for humans and mosquitoes, respectively.
\subsection{Finite-Difference Schemes}
We develop a finite-difference scheme for the proposed age-structured model with immunity feedback. A suitable numerical scheme should mimic the biological properties of the system, including positivity preservation and the conversation laws of population size and immunity. Our numerical scheme achieves these properties, while not imposing severe time-step constraints.

We consider a uniform grid with nodes $\alpha_k$, $t_n$ in the age and time dimensions, such that $\alpha_{k+1}-\alpha_{k} = \da$ and $t_{n+1}-t_{n} = \dt,~\forall k,n$, and we let $\da = \dt.$ The approximated value of quantity $Q$ at any point $(\alpha_{k},t_{n})$ on the discretized domain is denoted by $Q(\alpha_{k},t_{n})\approx Q^\kn$. Following the general idea in \cite{li2020age}, we employ an implicit-explicit approach and derive the following scheme:
\begin{subequations}\label{eq.num1}
\begin{align}
S_H^\knp &= \frac{S_H^\kn+\dt(\phi^\kn r_D D_H^\kn+r_A A_H^\kn+w V_H^\kn)}{1+\dt(\Lambda_H^n+\edit{\eta^{k+1}}\,\nu_p^\knp+\mu_H^\kp)},\\
E_H^\knp &= \frac{E_H^\kn+\dt\Lambda_H^n S_H^\knp}{1+\dt(h+\mu_H^\kp)},\\
A_H^\knp &\!=\! \frac{(1-r_A\dt)A_H^\kn\!+\!\dt\left((1-\rho^\kn)h E_H^\knp\!+\!(1-\phi^\kn)r_D D_H^\kn\right)}{1+\dt(\psi^\kn \Lambda_H^n+\mu_H^\kp)}\label{eq.num1_A},\\
D_H^\knp &=\frac{(1-r_D\dt)D_H^\kn+\dt\left(\rho^\kn h E_H^\knp+ \psi^\kn\Lambda_H^n A_H^\knp\right)}{1+\dt(\mu_H^\kp+\mu_D^\kp)},\label{eq.num1_D}\\
V_H^\knp &= \frac{(1-w \dt)V_H^\kn+\dt\,\edit{\eta^\kp}\,\nu_p^\knp S_H^\knp}{1+\dt\,\mu_H^\kp},\label{eq.num1_V}
\end{align}
\end{subequations}
where 
$\Lambda_H^n=b_H\big(N_M^n,N_H^n\big)\, \beta_M\, I_M^{\star,n}/N_M^n$, $\phi^\kn=\phi(C_H^\kn/P_H^\kn)$, $\rho^\kn=\rho(C_H^\kn/P_H^\kn)$, and $\psi^\kn=\psi(C_H^\kn/P_H^\kn).$ 
\edit{The boundary conditions for $S_H(0,t)$ in \cref{eq.BC_vh} and $C_m(0,t)$ in \cref{eq:immune_BC} are discretized using the trapezoidal rule.}

For the sake of efficiency, we have approximated the time derivatives using backward Euler and impose implicit discretization (at time level $t_{n+1}$) only when the values are available to avoid solving additional linear systems. 

Denoting \edit{$P_H^{k,n} = S_H^{k,n}+E_H^{k,n}+A_H^{k,n}+D_H^{k,n}+V_H^{k,n}$}, the discretization \cref{eq.num1} satisfies
\begin{equation*}
\edit{\frac{P_H^\knp-P_H^{k,n}}{\dt}} = -\mu_H^\kp P_H^\knp - \mu_D^\kp D_H^\knp.
\end{equation*}
Hence all the stage progression terms are balanced with each other at each time step, and the resulting scheme preserves the conservation law of population. When $\mu_D=0$, this is consistent with \cref{eq.P_H_PDE}. From \cref{eq.num1_A,eq.num1_D,eq.num1_V}, we obtain the time-stepping constraint for maintaining the positivity of the population size
\begin{equation*}
1-r_A\dt\ge 0,\quad  1-r_D\dt\ge 0,\quad\text{and}\quad  1-w\dt\ge 0,
\end{equation*}
which is not restrictive in practice, given the baseline values of  $r_A$, $r_D$, and $w$. 

Similarly, we discretize the immunity system \cref{eq.basic_model_immunity} by evaluating all the terms on the right-hand side at $(\alpha_{\kp},t_{n+1})$, and we obtain
\begin{align*}
C_e^\knp & = \frac{C_e^\kn+\dt B^\knp}{1+\dt\,(1/d_e+M^\knp)},\quad 
C_m^\knp = \frac{C_m^\kn}{1+\dt\,(1/d_m+M^\knp)},\\
C_{\nu}^\knp & = \frac{C_{\nu}^\kn+\dt\,c_{\nu} \,\nu_b^\knp S_H^\knp }{1+\dt\,(1/d_{\nu}+M^\knp) }, \quad \text{where}
\end{align*}  
\[
B^\knp = f(\Lambda_H^{n+1})\left(c_S S_H^\knp + c_E E_H^\knp + c_A A_H^\knp + c_D D_H^\knp\right),
\] 
and $M^\knp = \mu_H^\kp+\mu_D^\kp D_H^\knp /P_H^\knp$.
\subsection{Model Calibration} \label{sec:cali}
We parametrize our model using demographic and immunological data from areas in sub-Saharan Africa. 
\paragraph{Calibration of Demographic Structure}
With Kenya as our baseline population, we employ a scaled skew normal distribution \cite{mazzuco2015fitting} for the fertility rate function, 
\[
g_H(\alpha) =\frac{2\,b_4}{b_1} \varphi\Big( \frac{\alpha/365-b_2}{b_1} \Big)\Phi\bigg(b_3\Big(\frac{\alpha/365 - b_2}{b_1}\Big) \bigg)/(365\times2),
\]
where $\varphi(\alpha)$ and $\Phi(\alpha)$ are the probability density function and cumulative distribution function for the standard normal distribution, respectively. The fertility per person is assumed to be half of the fertility per woman. We fit the model to the data in \cite[Table 5.1]{kenyanationalbureauofstatistics2015kenya} and obtain coefficients $b_1 = 13.20$, $b_2 = 17.96$, $b_3 = 4.08$, and $b_4 = 4.02$.

\begin{figure}[htbp]
\centering
\subfloat[Demographic curves]{\includegraphics[width=0.33\textwidth]{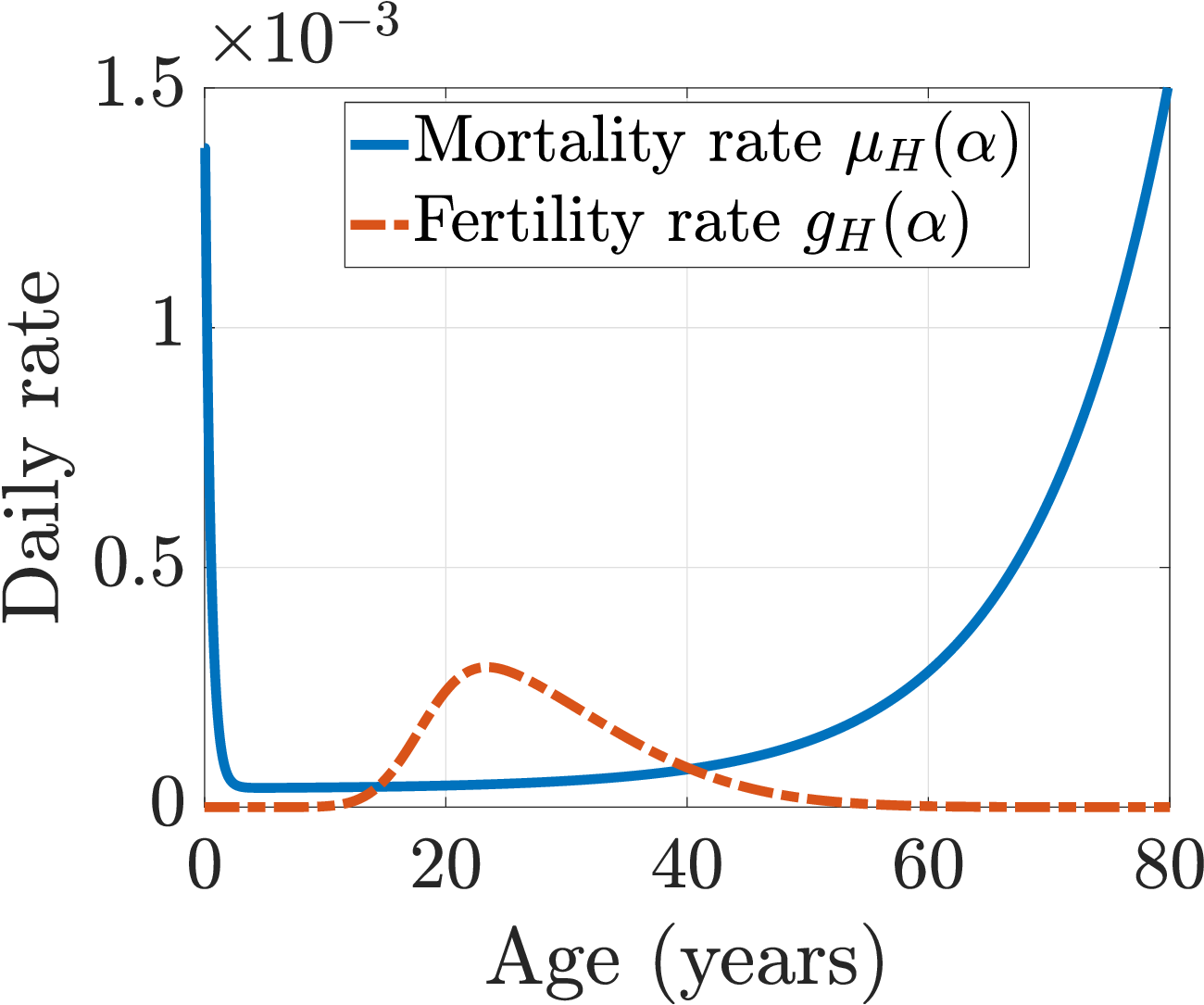}\label{fig:demo_a}}
\hfill\subfloat[Stable age distribution]{\includegraphics[width=0.33\textwidth]{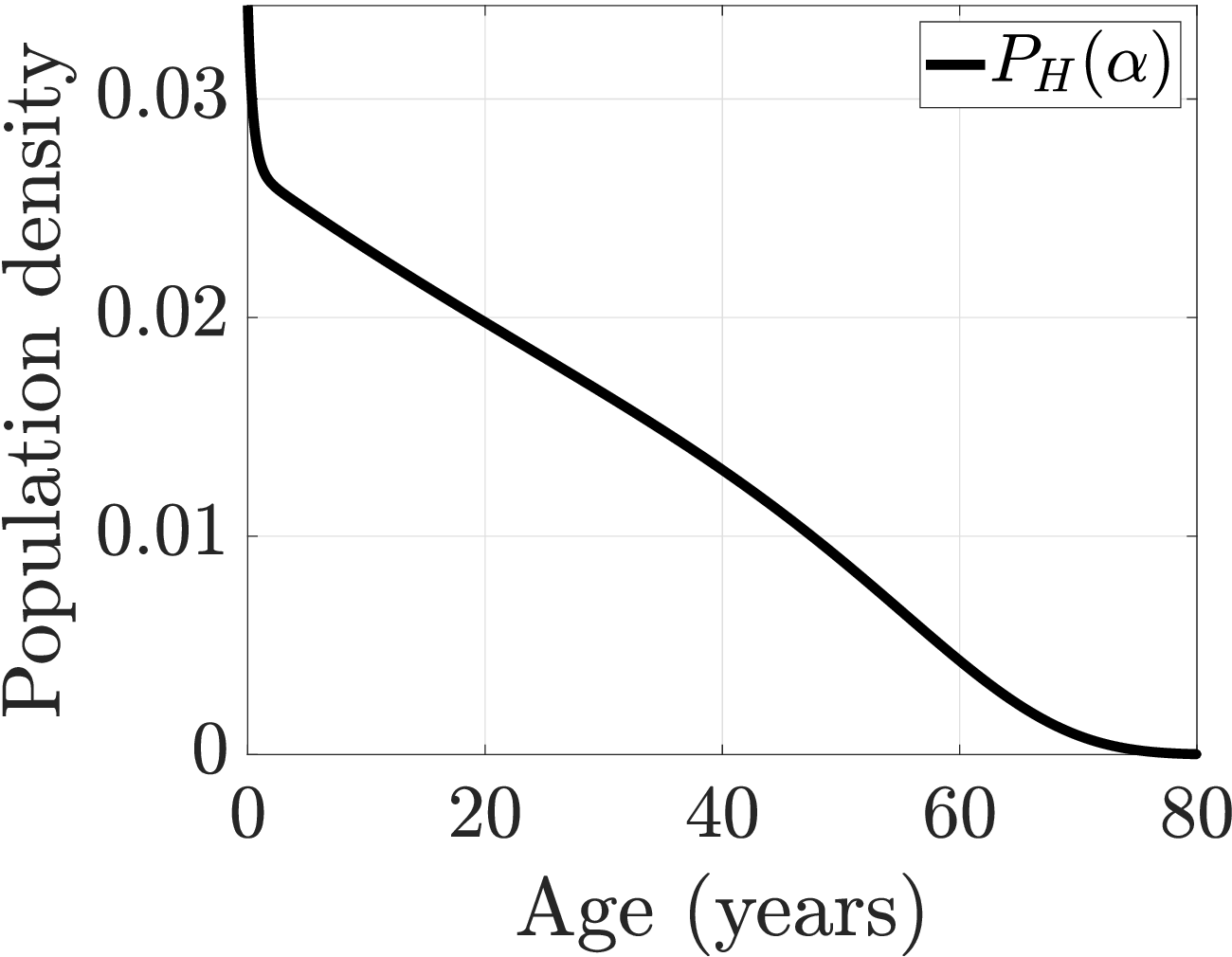}\label{fig:demo_b}}\hfill\subfloat[Linking functions]{\includegraphics[width=0.33\textwidth]{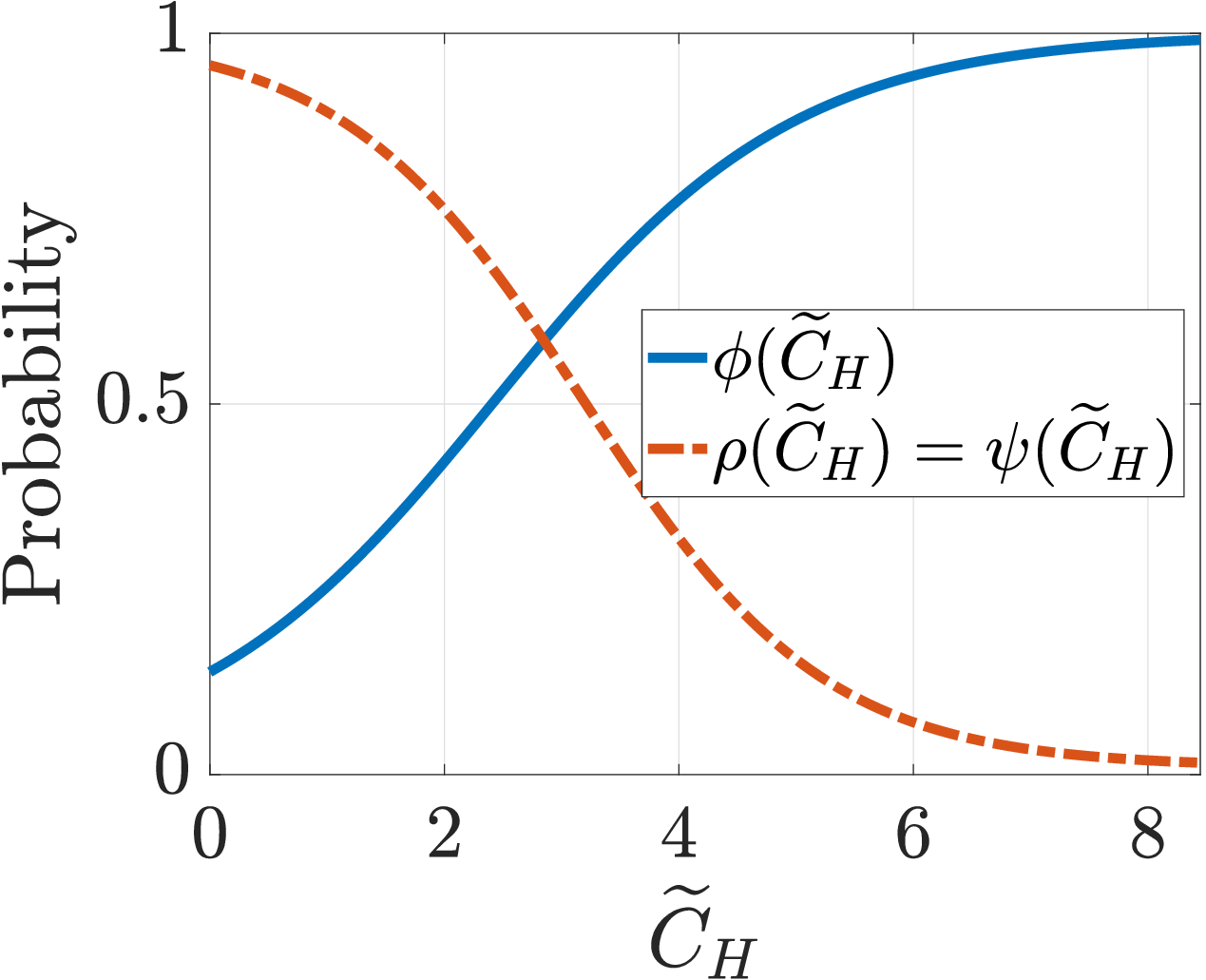}\label{fig:sigmoid}}
\caption{Model calibration results as described in \cref{sec:cali}.}
\end{figure}

To have a population with a constant demographic structure, we calibrate the natural mortality rate $\mu_H(\alpha)$ by setting $q=0$ in \cref{eq.q_zero} and solving for a balanced mortality distribution $\mu_H(\alpha)$. We fit a three-component competing-risk model \cite{siler1979competing}, 
\[
\mu_H^{(0)}(\alpha) = \Big(d_1+ d_2 e^{-d_3 \alpha/365}+d_4 e^{d_5\alpha/365}\Big)/365,
\]
using the mortality estimates from \cite{WHO_data}, where $d_1 = 0.002$, $d_2 = 0.09$, $d_3 = 2.1$, $d_4 = 10^{-4}$, $d_5 = 0.09$. The balanced mortality rate $\mu_H(\alpha) \approx 5.8\, \mu_H^{(0)}(\alpha).$ The calibrated fertility and mortality and the corresponding stable population age distribution are in \cref{fig:demo_a,fig:demo_b}, respectively.

\paragraph{Calibration of Immunity Linking Parameters}
We calibrate our model to the immunity curves in \cite[Figure 7B]{filipe2007determination}, where the population's susceptibility to developing clinical disease (corresponding to $\rho$ and $\psi$) is a function of age and environmental exposure levels, measured by the annual entomological inoculation rate (aEIR), $
aEIR=b_H \frac{I_M^\star}{N_M}$. We study medium to high aEIR ($\ge20$) since low aEIR corresponds to regions with sporadic malaria incidence, which are not the focus of this paper.

To describe the immunity feedback onto disease transmission through the immunity-dependent probabilities $\rho,\ \phi,\ \psi$, we pick a sigmoid-shaped linking function 
\begin{equation}
\mathcal{S}(x;f_0,f_1, s,r) = f_0+\frac{f_1-f_0}{1+e^{-(x-s)/r}},    
\label{eq.link}
\end{equation}
and assume that responses in the progression probabilities to severe disease are similar, that is, $\phi(\widetilde{C}_H) = \mathcal{S}(\widetilde{C}_H;f_0,f_1, s_0,r_0)$, $\rho(\widetilde{C}_H) = \psi(\widetilde{C}_H) = \mathcal{S}(\widetilde{C}_H;f_1,f_0, s_1,r_1)$. We fix $f_0=0.01, f_1 = 1$, and the calibrated parameters $s_0 = 2.43, r_0 = 1.28, s_1 = 3.19, r_1 = 1.03$, and the corresponding curves are plotted in \cref{fig:sigmoid}.

\begin{figure}[htbp]
\centering
\subfloat[]{\includegraphics[width=0.33\textwidth]{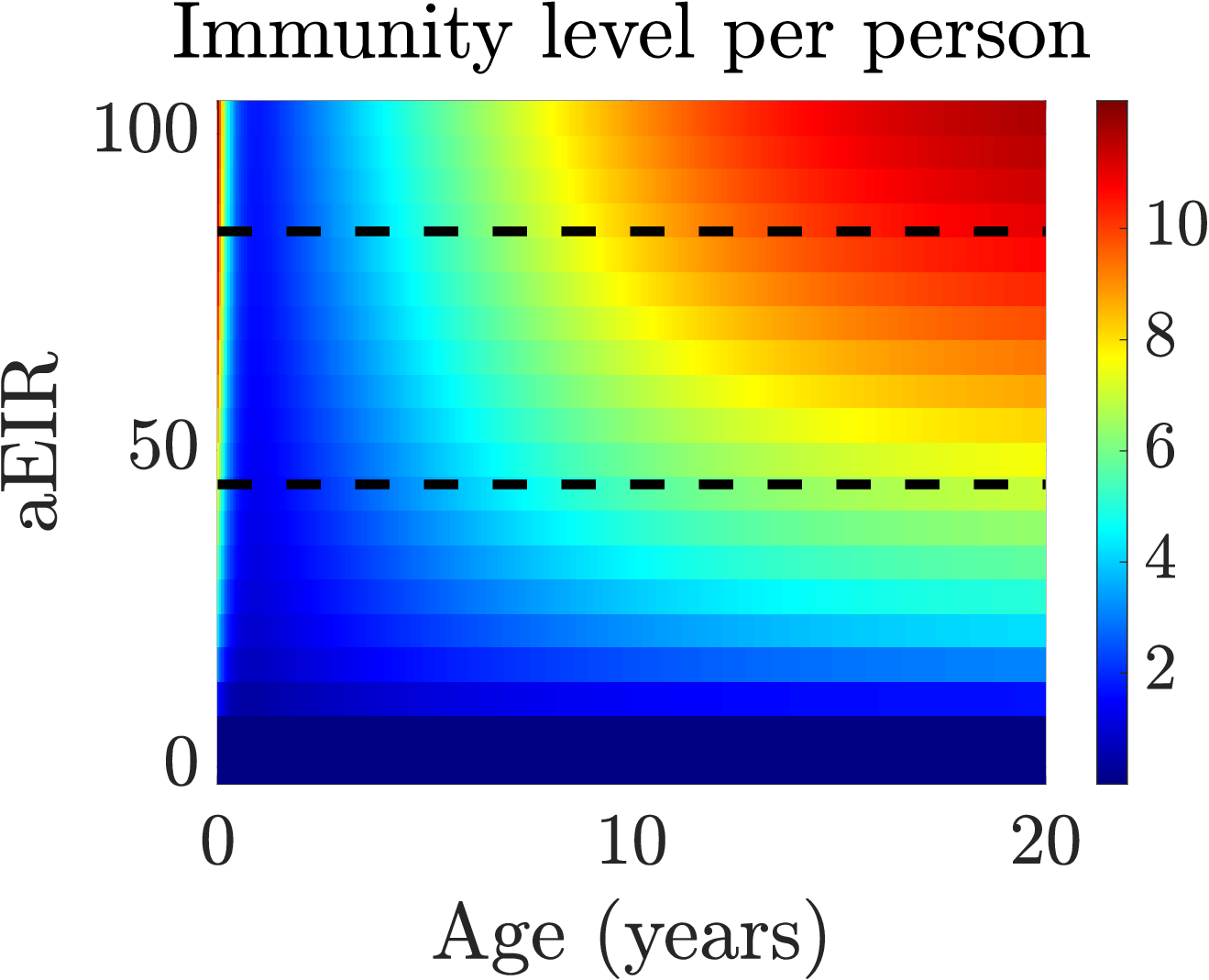}\label{fig:heatmap_a}}\hfill
\subfloat[aEIR = 44.66]{\includegraphics[width=0.33\textwidth]{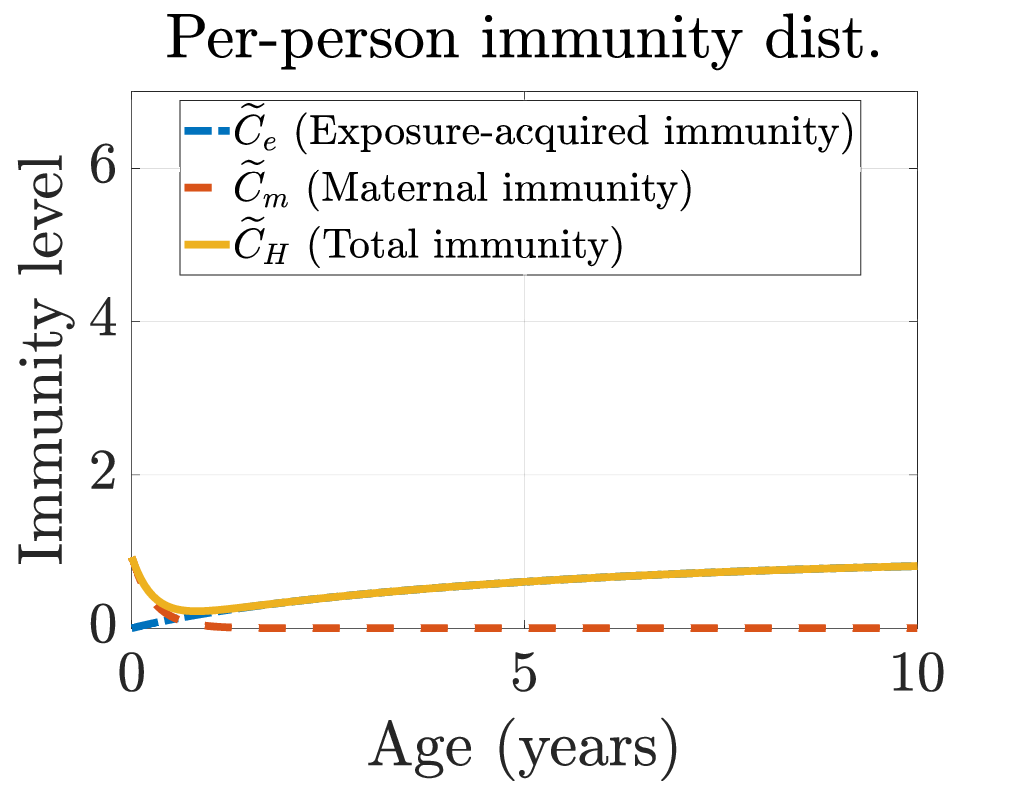}\label{fig:heatmap_b}}\hfill
\subfloat[aEIR = 84.61]{\includegraphics[width=0.33\textwidth]{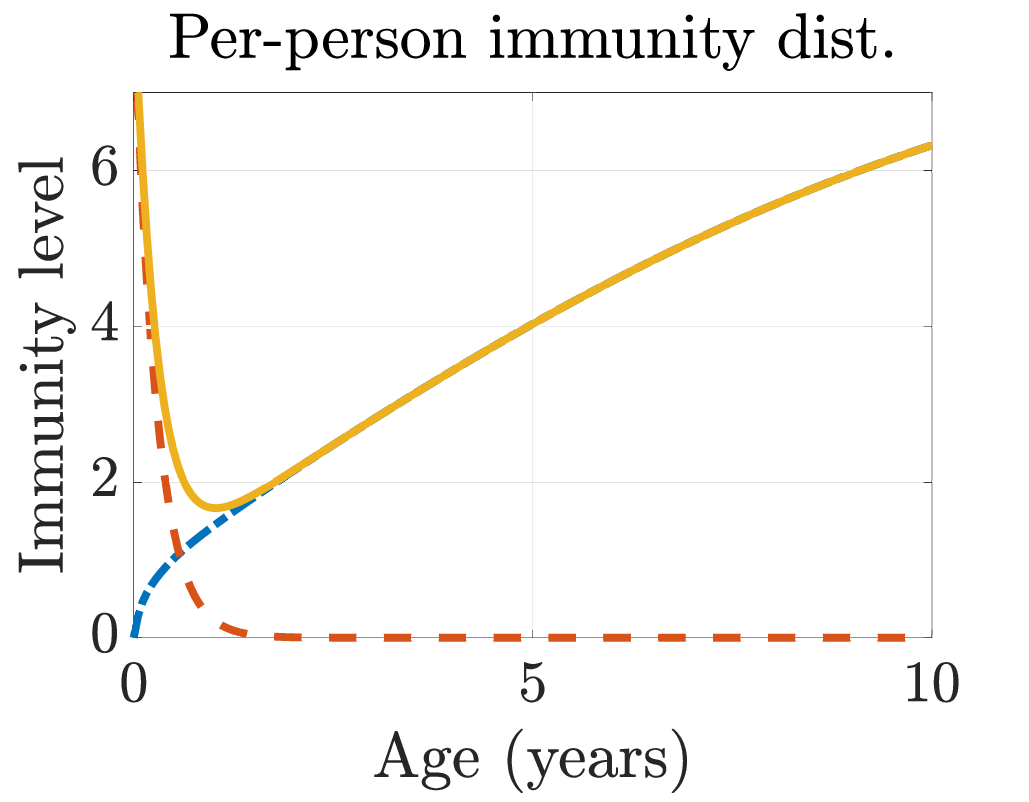}\label{fig:heatmap_c}}
\caption{(a) Per-person immunity profile, $\widetilde{C}_H$, in the population depends on the age ($x$-axis) and the exposure level ($y$-axis). The cross sectional plots are: (b) for a low-transmission region ($\beta_M = 0.008$, $\mathcal{R}_0=1.24$) and (c) for a high-transmission region ($\beta_M = 0.25$, $\mathcal{R}_0=6.93$).} 
\end{figure}

\paragraph{Calibrated Baseline Scenario}
We calibrate the model relative to a baseline scenario with $\mathcal{R}_0 \approx 6.93$ (shown in \cref{fig:heatmap_c}) and aEIR $\approx 84.61$. \Cref{fig:heatmap_a} shows the distribution of per-person immunity profile in different transmission settings; we vary $\beta_M$ (mosquito infectivity) to create a range of aEIR and keep all the other parameters at baseline values. The maternal immunity level decays quickly after the birth for all the aEIR values. In the lower transmission setting (\cref{fig:heatmap_b}), the exposure-acquired immunity profile is flat across all ages, while in the high-transmission setting (\cref{fig:heatmap_c}), exposure-acquired immunity is boosted as people get older and receive repeated exposure. Filipe et al. \cite{filipe2007determination} required both anti-disease and anti-parasite immunity to see profiles by age that differed in low and high transmission settings. In contrast, our model only requires dynamic immunity feedback with anti-disease immunity to see the variation in immunity profiles as transmission settings differ.

\subsection{Impact of Immunity Feedback}
\begin{figure}[htbp]
\centering
\subfloat[]{\includegraphics[width=0.325\textwidth]{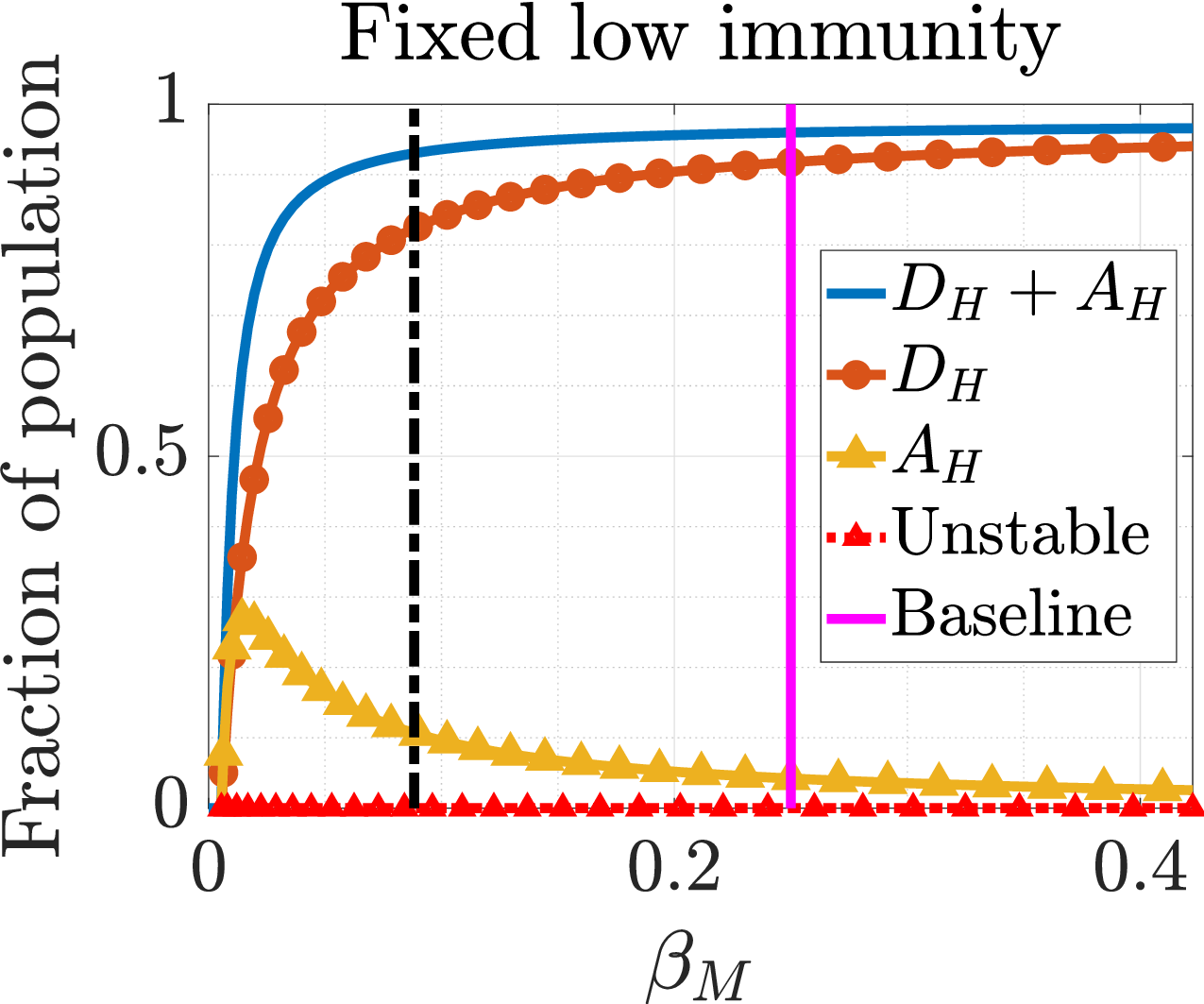}\label{fig:bifur_a}}\hfill\subfloat[]{\includegraphics[width=0.325\textwidth]{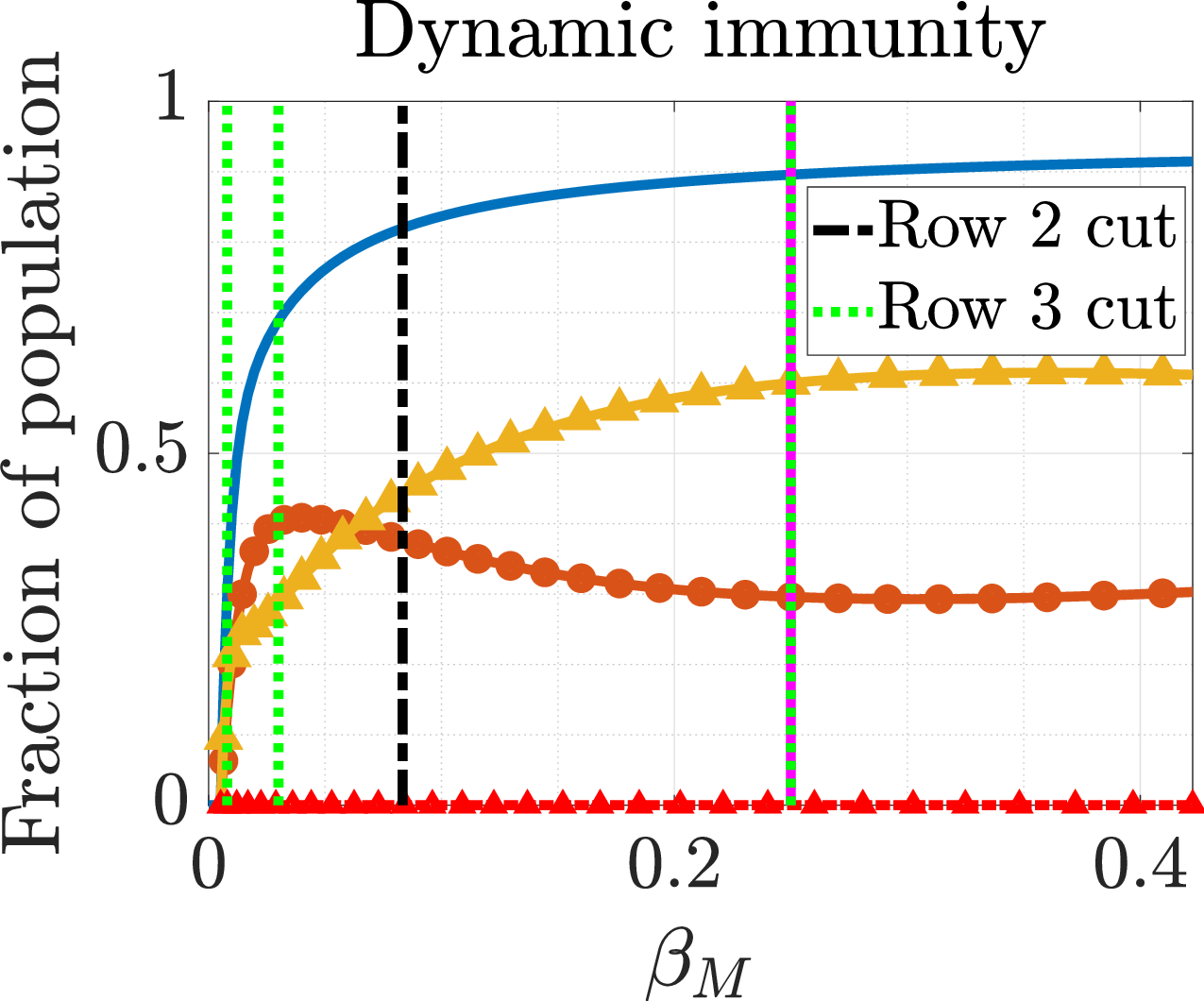}\label{fig:bifur_b}}\hfill\subfloat[]{\includegraphics[width=0.325\textwidth]{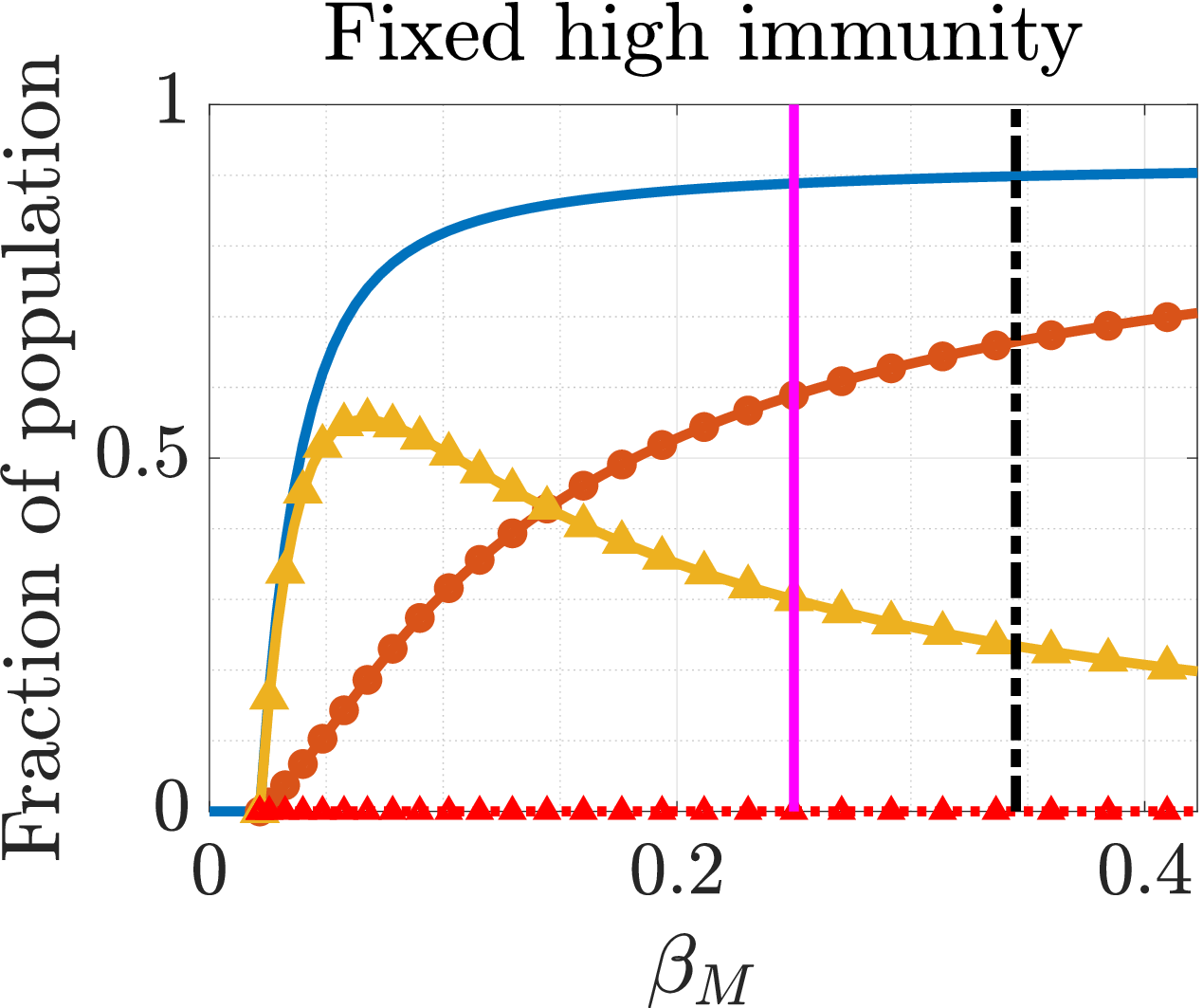}\label{fig:bifur_c}}\\ \vspace{-0.3cm}
\subfloat[]{\includegraphics[width=0.325\textwidth]{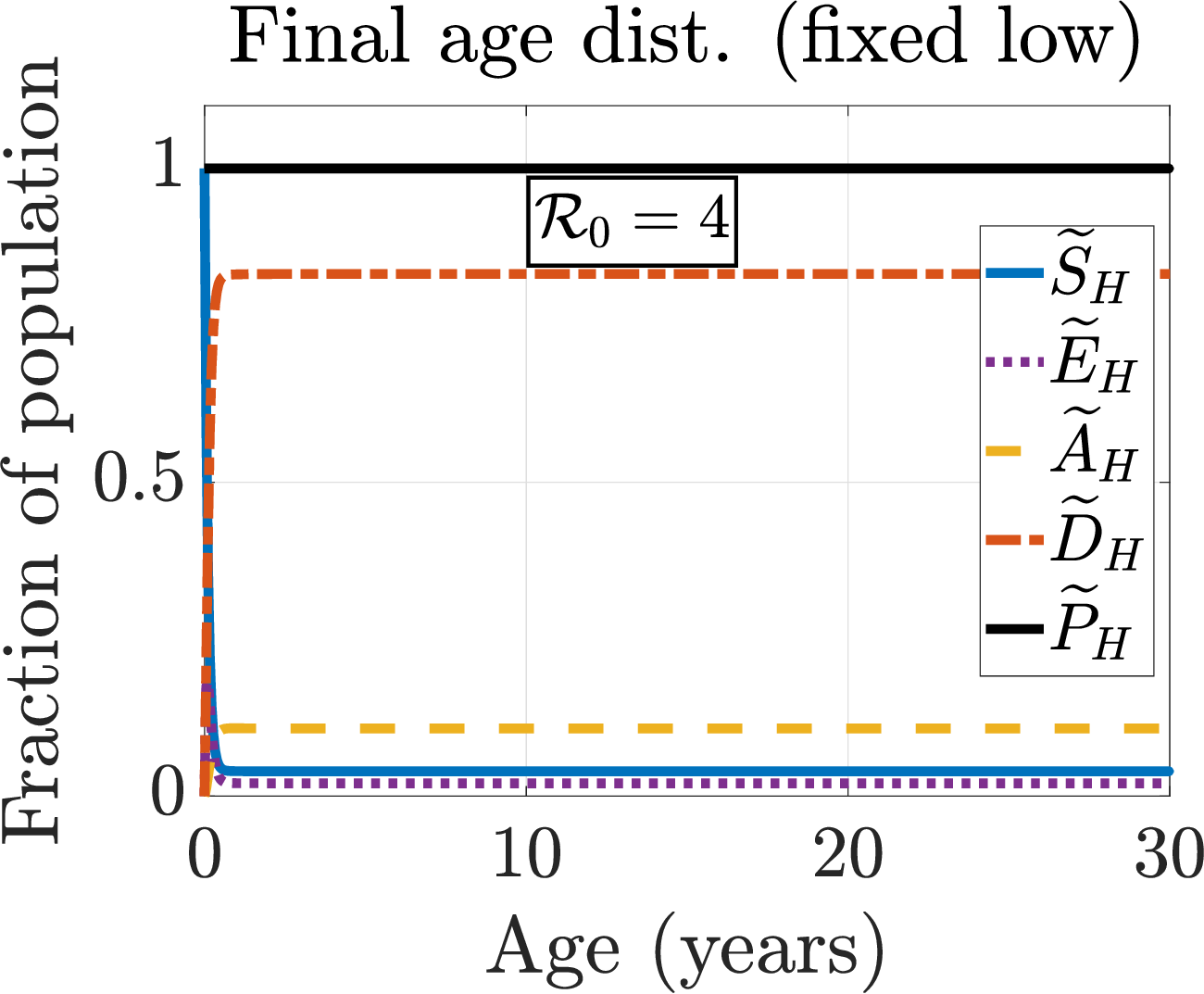}\label{fig:bifur_d}}\hfill\subfloat[]{\includegraphics[width=0.325\textwidth]{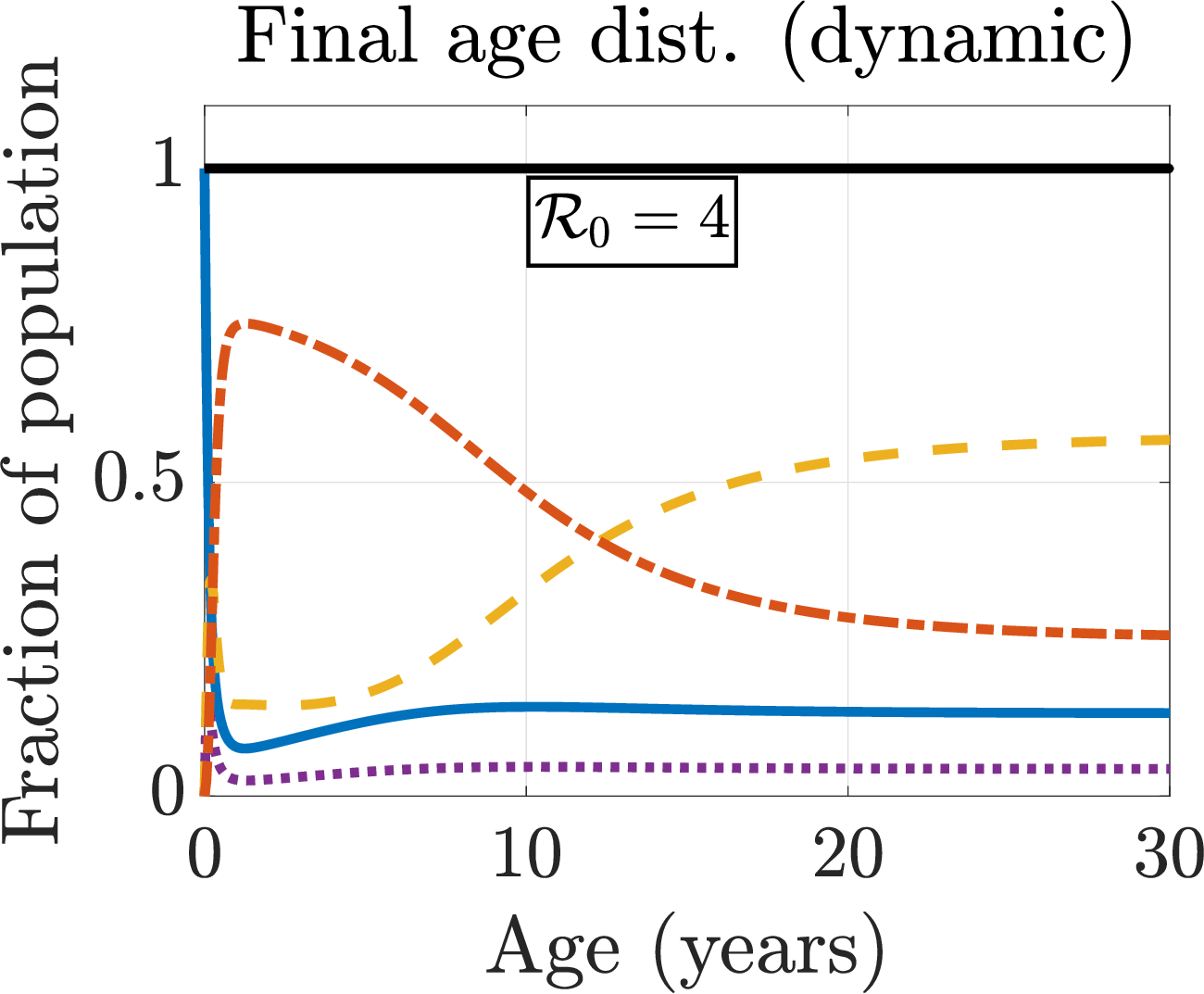}\label{fig:bifur_e}}\hfill\subfloat[]{\includegraphics[width=0.325\textwidth]{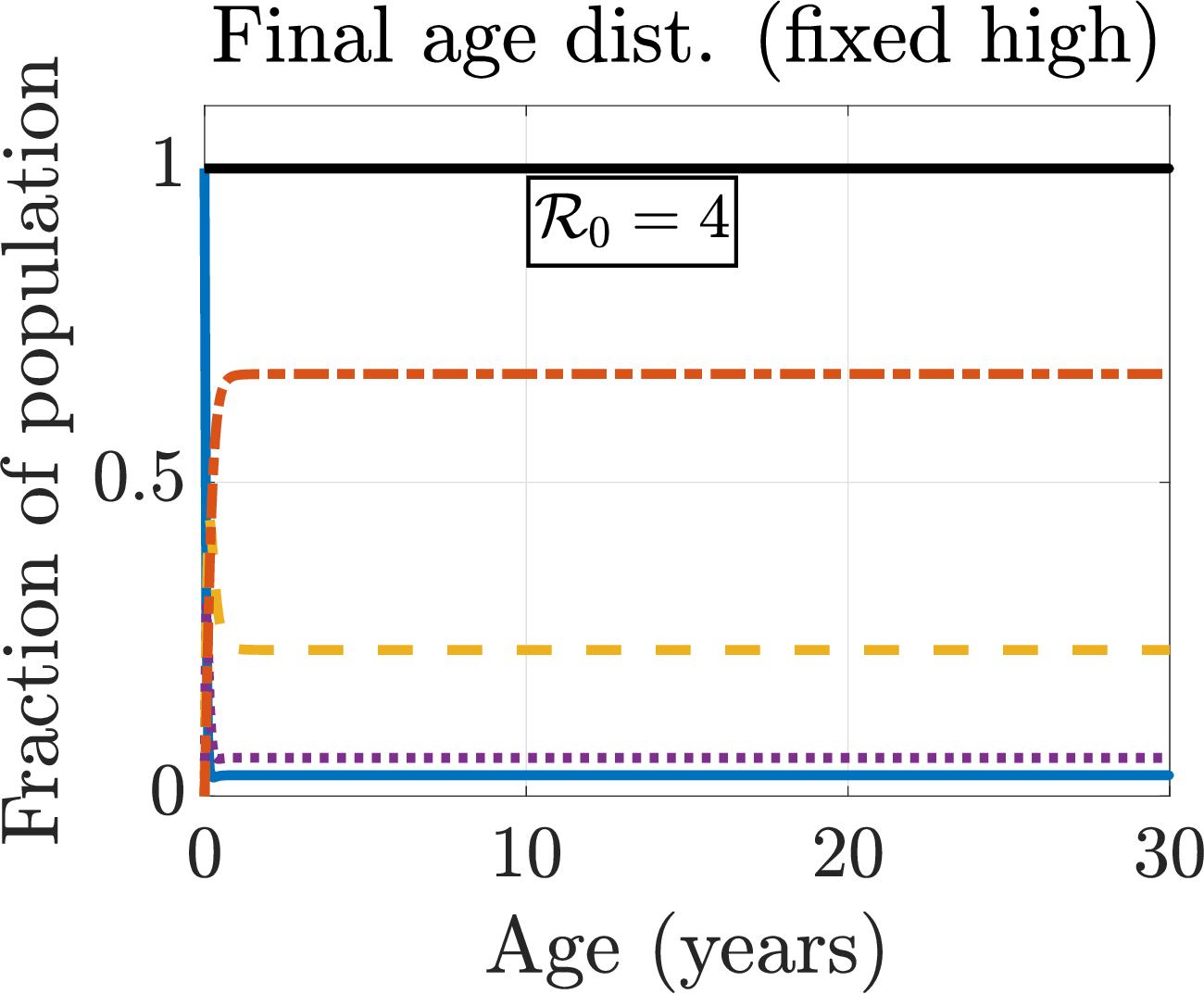}\label{fig:bifur_f}}\\ \vspace{-0.3cm}
\subfloat[]{\includegraphics[width=0.325\textwidth]{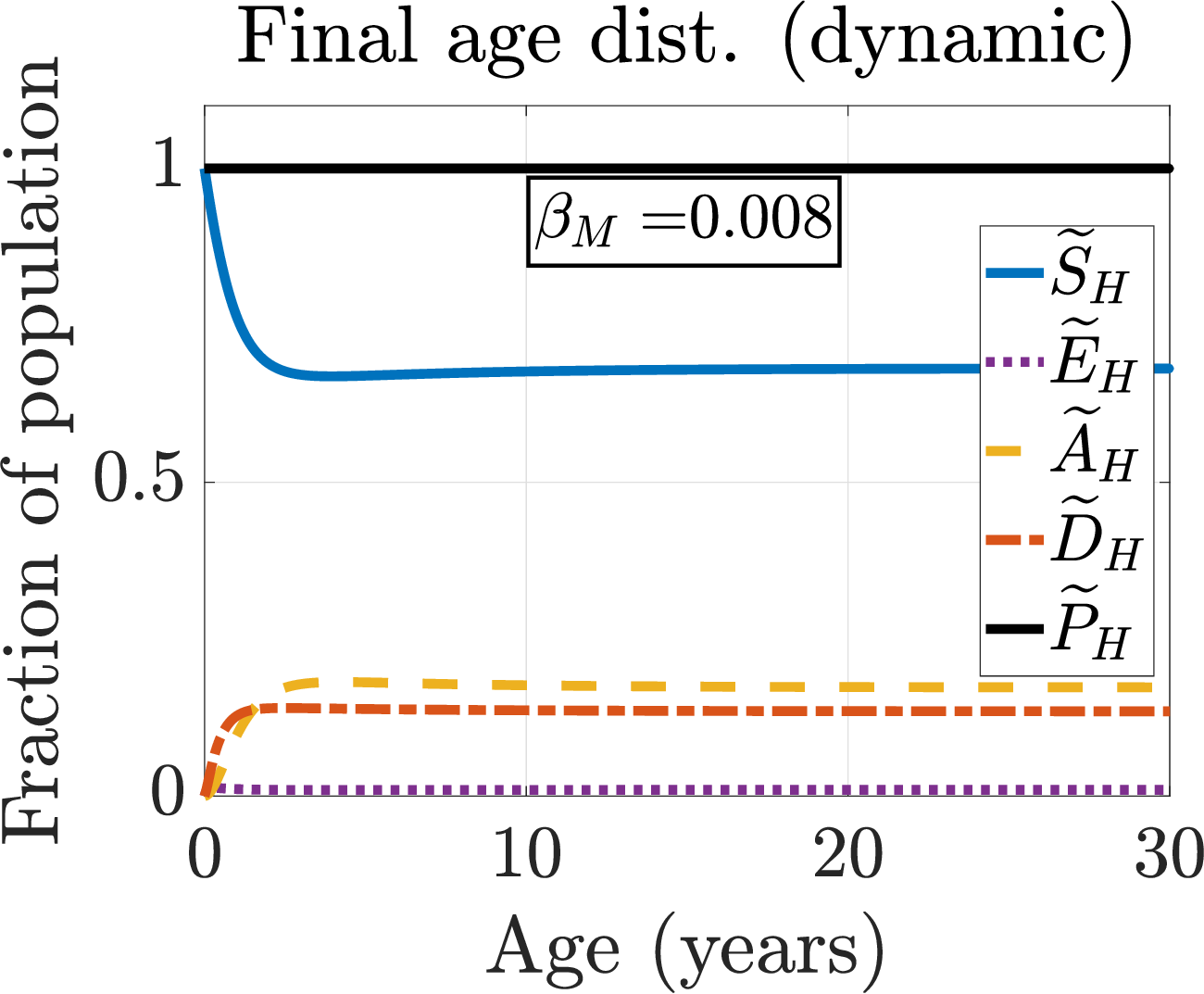}\label{fig:bifur_g}}\hfill\subfloat[]{\includegraphics[width=0.325\textwidth]{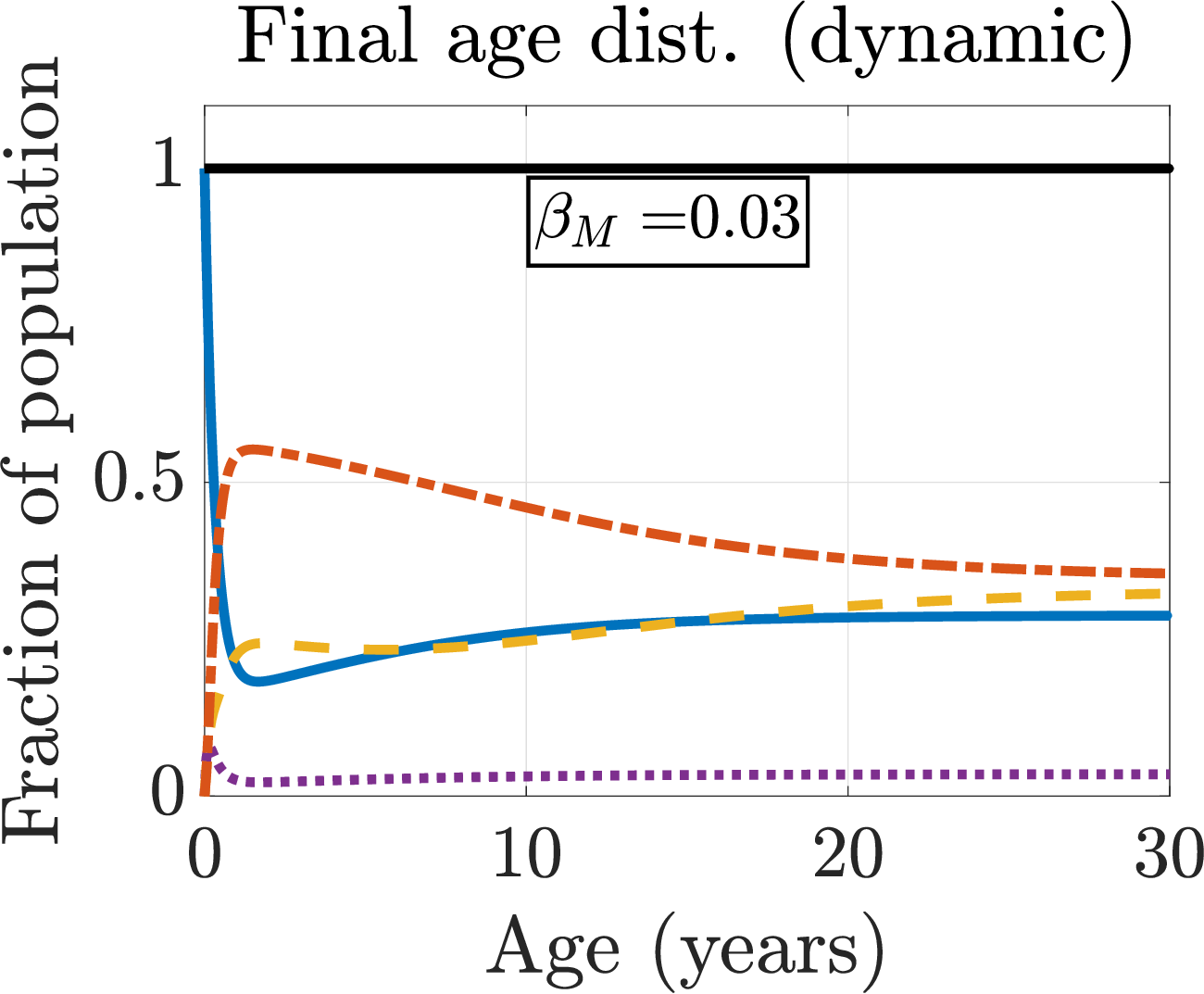}\label{fig:bifur_h}}\hfill\subfloat[]{\includegraphics[width=0.325\textwidth]{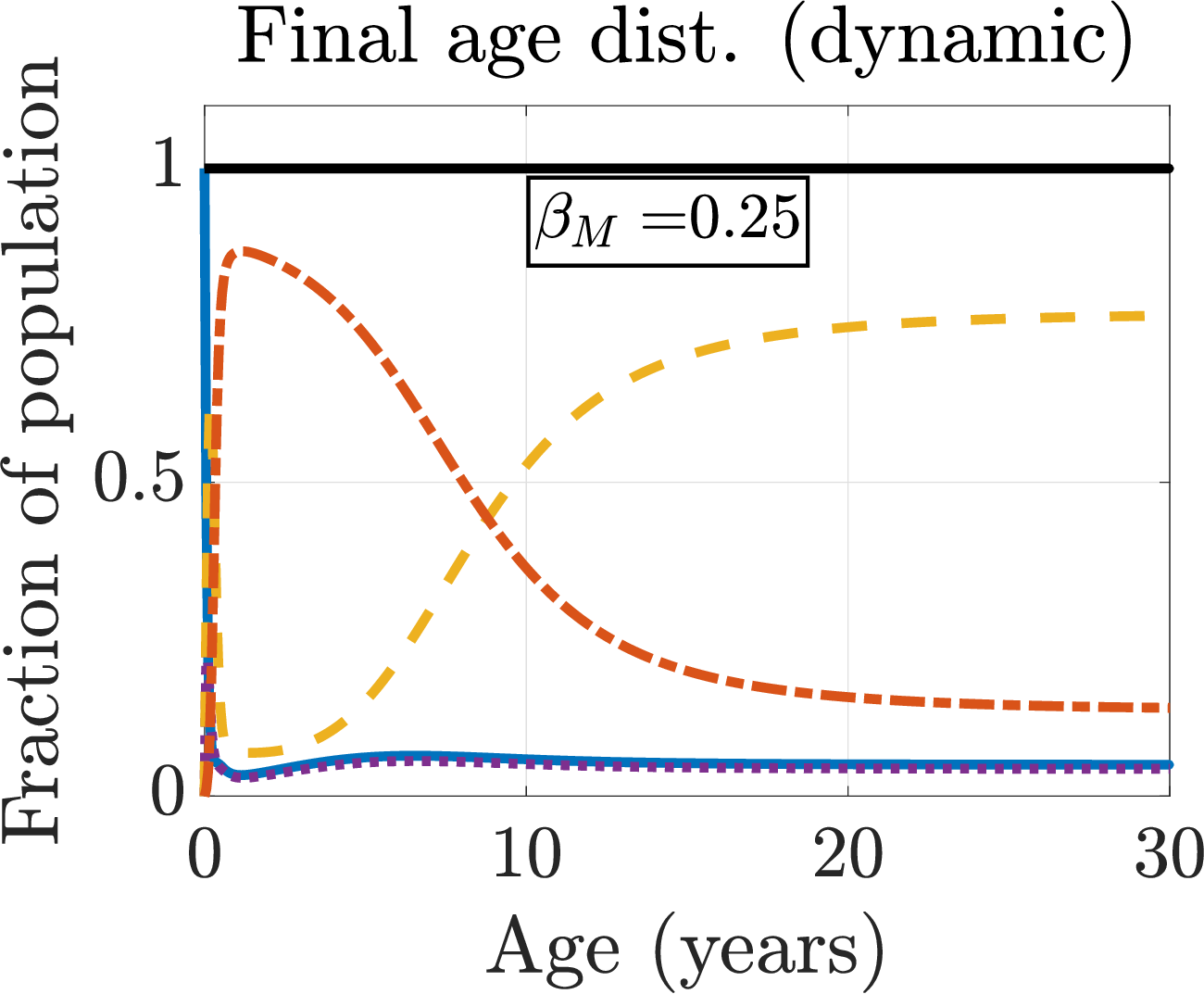}\label{fig:bifur_i}}
\caption{\textbf{Row 1}: Comparison of bifurcation diagrams with fixed low immunity, dynamic immunity, and fixed high immunity. \Cref{fig:supp_fig_3} shows the bifurcation plots using $\mathcal{R}_0$ as $x$-axis, and  \cref{fig:supp_fig_1} shows the age-structure of the endemic equilibrium across the range of $\beta_M$ in each case. \textbf{Row 2}: Age-distributions of infection status at endemic equilibrium when $\mathcal{R}_0=4$ in each immunity set up (using $\beta_M$ values indicated by the vertical dash-dot lines in black). \textbf{Row 3}: Age-distributions of infection status at endemic equilibrium for dynamic immunity case (using $\beta_M$ values indicated by the vertical dotted lines in green)  \label{fig:comp_bifur}}
\end{figure}

\paragraph{Impact on Endemic Equilibrium and Bifurcation} We numerically capture the stable endemic equilibrium and obtain a forward bifurcation using parameter $\beta_M$ in \cref{fig:bifur_b} (\cref{fig:supp_fig_3} gives the plots using $\mathcal{R}_0$ in the $x$-axis). As $\beta_M$ (exposure level) increases, the fraction of asymptomatic population $A_H$ keeps increasing, and the fraction of severe disease population, $D_H$, starts decreasing when $\beta_M>0.03  ~(\mathcal{R}_0>2.5)$. This reflects the fact that, due to a stronger exposure, dynamic immunity creates larger feedback on the progression parameters, which end up reducing the fraction of severe disease among the infectious groups $D_H/(A_H+D_H)$. We also notice that the $D_H$ curve hits a local minimal around $\beta_M = 0.3~(\mathcal{R}_0 = 7.6)$ with immunity feedback. 

We then plot the bifurcation diagram with immunity feedback turned off, where we assume constant disease progression parameters regardless of the change in the population immunity level. For example, we take
\[
\bar{\rho}(\beta_M) = \frac{1}{N_H}\edit{\int_0^A} \rho\big(\widetilde{C}_H^{EE}(\alpha;\beta_M)\big) \edit{P_H^*(\alpha)} d\alpha,
\]
where $\widetilde{C}_H^{EE}$ is the total per-person immunity level at the endemic equilibrium, which varies depending on exposure level, parametrized by $\beta_M$. The values for $\bar{\phi}$ and $\bar{\psi}$ are defined similarly. The calculated  $\bar{\rho}$ gives the population average of the transition parameter $\rho(\widetilde{C}_H)$ at the endemic equilibrium (under dynamic immunity); this allows a more fair comparison with constant immunity settings. We consider a population with a fixed low-immunity profile ($\beta_M=0.008$), which gives $\bar{\rho} =\bar{\psi}=0.91,~\bar{\phi} = 0.23$. In addition, we consider a fixed high-immunity scenario with $\beta_M = 0.25$, which gives $\bar{\rho} =\bar{\psi}=0.11,$ and $\bar{\phi} = 0.92$. 

As shown in \cref{fig:bifur_a,fig:bifur_c}, at higher exposure levels, without dynamic immune feedback, the fraction of severe disease increases monotonically, and the fixed low-immunity case gives worse disease outcomes. Compared with the dynamic-immunity case (\cref{fig:bifur_b}), under the same exposure level ($\mathcal{R}_0=4$, $\beta_M$ at vertical dash-dot lines in black), there are more infectious people ($A_H+D_H$) for the constant-immunity cases, and the fraction of asymptomatic individuals is much lower. 

\paragraph{Impact on Age Distribution of the Disease Status}
Dynamic immunity creates heterogeneity in the age-distributions of infection curves (\cref{fig:bifur_e}), while the fixed-immunity cases give almost homogeneous distributions (\cref{fig:bifur_d,fig:bifur_f}), except a fast transition near age zero due to susceptibility following maternal protection. 

The amount of heterogeneity in the age distribution, under the dynamic-immunity setting, also depends on the exposure level. When there is low malaria transmission (\cref{fig:bifur_g}), the boosting in the exposure-acquired immunity is limited, and there is almost no heterogeneity in the age distribution curves beyond three years of age. As the transmission level increases (\cref{fig:bifur_h,fig:bifur_i}), more heterogeneity is developed among young children under 15 years of age. At the baseline scenario (\cref{fig:bifur_i}), the fraction of severe disease peaks around one year old. Once the immunity level achieves a sufficiently high level through repeated exposure, the fraction of symptomatic infection settles down to a relatively low constant level. Moreover, given the large young population cohort in the current demographic setting (\cref{fig:demo_b}), children in this age range have the highest severe disease counts.

These observations, including decreases of disease severity with age, qualitatively match field observations from high-transmission regions \cite{filipe2007determination,rodriguez2018quantification,smith2007standardizing}. Thus, our results confirm the necessity of employing a dynamic mechanism in tracking the population immunity and linking it with epidemiological parameters. 

\subsection{Exploration of Vaccination}
As a preliminary investigation, we simulate an RTS,S malaria vaccine implementation among young children using the vaccination rate $\nu_p(\alpha) = \nu_p^0 \,I_{\alpha \in [9\times 30, 10\times 30]}$, where $I_{(\cdot)}$ is the indicator function. This assumes that children complete a full three-dose series of RTS,S at around $9\sim10$ months old \cite{ministryofhealthofkenyakenya}. We consider a high daily per-capita vaccination rate for this age cohort, $\nu_p^0 =0.8$. Under this rate, among a population of about 9.4 million in the malaria endemic counties in Kenya, there are around 58,000 eligible children vaccinated per year (at the endemic equilibrium), which is within the capacity of the local infrastructure \cite{ministryofhealthofkenyakenya}.

\begin{figure}[htbp]
\centering
\subfloat[Stable age distribution]{\includegraphics[width=0.48\textwidth]{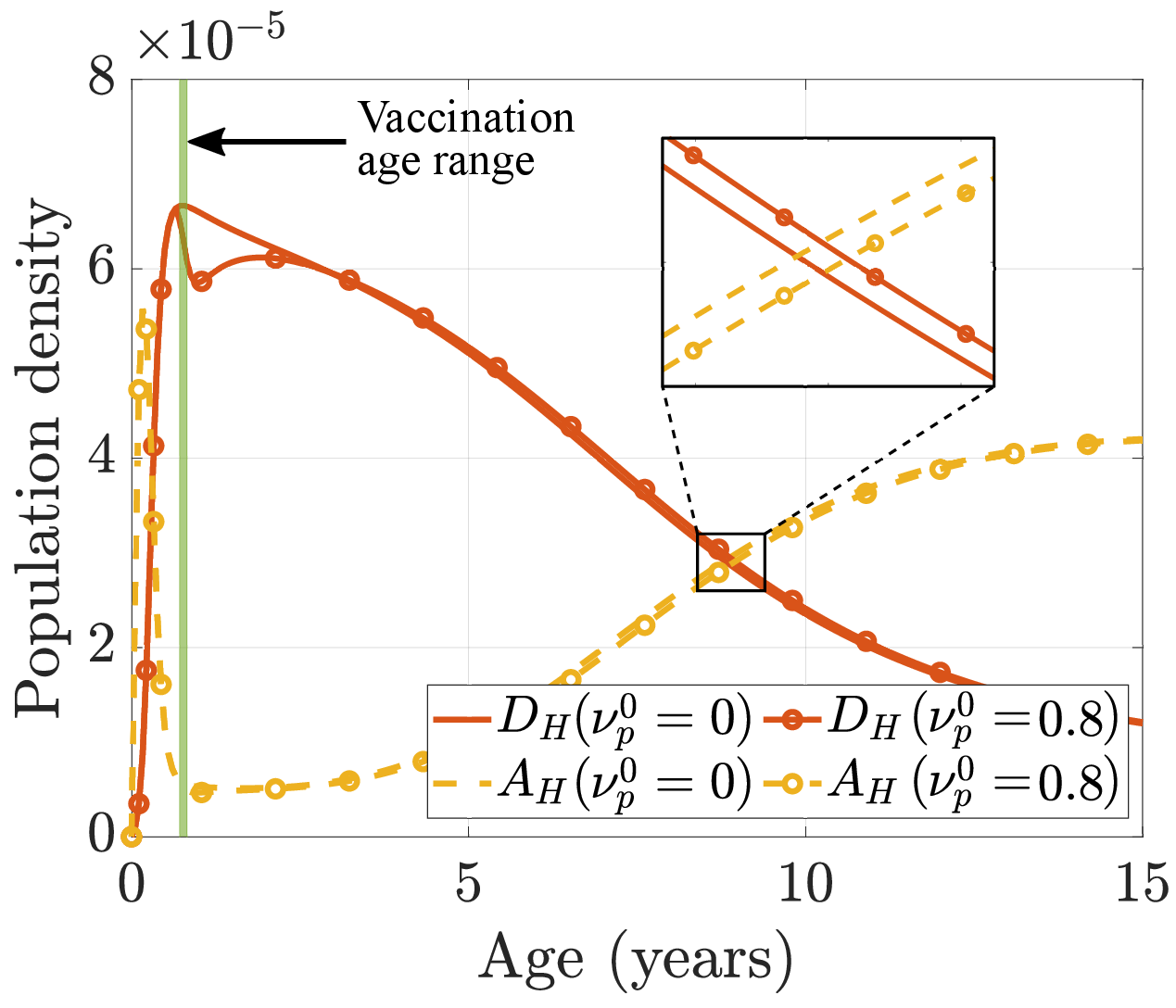}\label{fig:vaccine_a}}\hfill\subfloat[Per-person immunity distribution]{\includegraphics[width=0.48\textwidth]{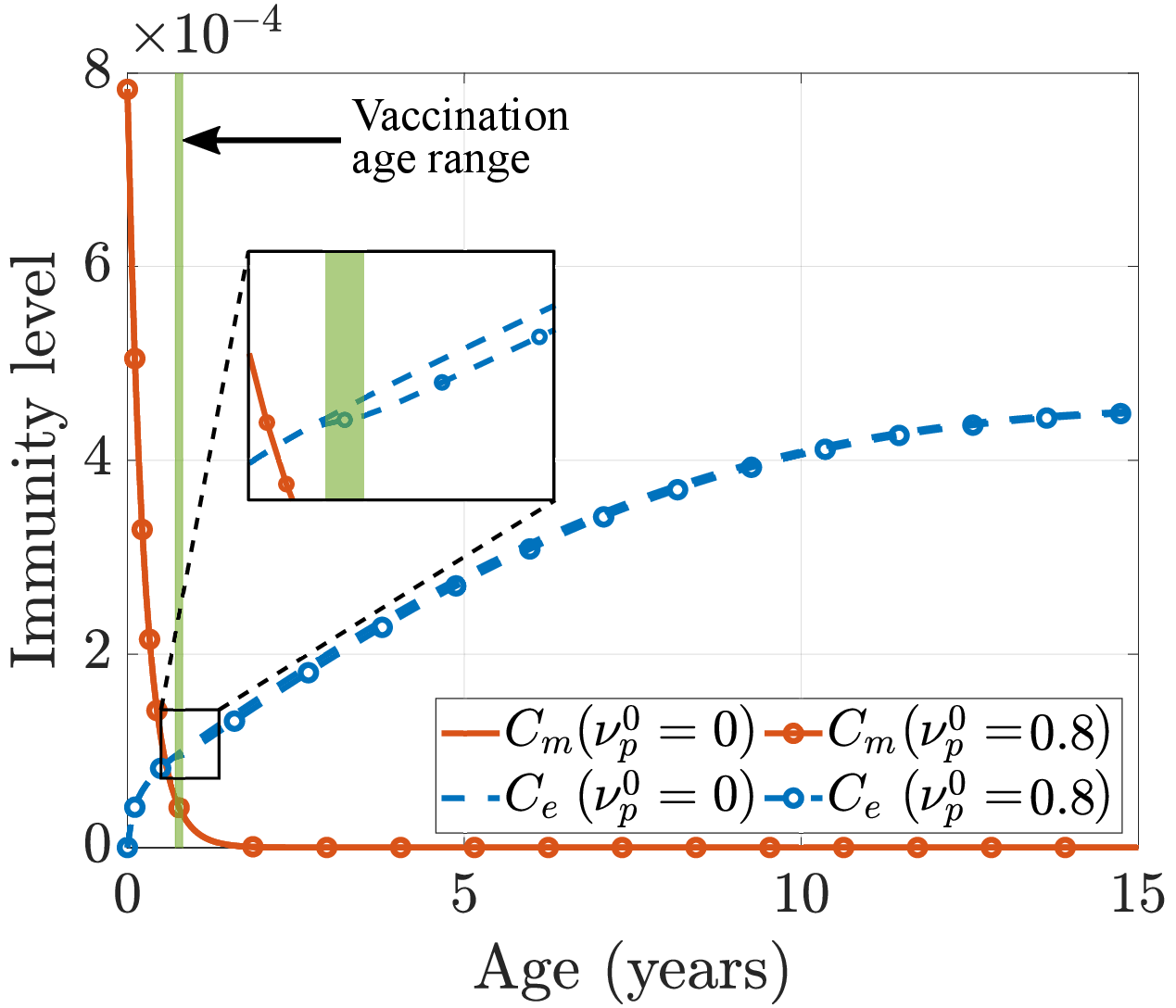}\label{fig:vaccine_b}}
\caption{Comparison of endemic equilibrium between vaccination level $\nu_p^0 = 0.8$ and no vaccination $\nu_p^0 = 0$. Vaccination lowers the severe disease cases before three years old and slightly increases that number for older ages (a), which results from the reduced exposure (b).\label{fig:vaccine}}
\end{figure}

\Cref{fig:vaccine} shows the impact under the prescribed vaccination setting. There is a drop in the number of severely diseased for children aged 9 months to about three years old (\cref{fig:vaccine_a}), which is about $20,716$ (or $4.05\%$ of) severe cases avoided. When above three years old, the vaccination leads to a slightly higher $D_H$ curve and a lower $A_H$ curve. This is due to the reduced exposure from the vaccination (\cref{fig:vaccine_b}), which results in a lower level of exposure-acquired immunity. Thus, vaccinated children are more likely to progress to $D_H$ stage rather than $A_H$ when being exposed at an older age. This effect has been reported in other malaria interventions that reduce exposure, where the age distribution of severe malaria peaks at older ages \cite{ceesay2008changes, griffin2010reducing,omeara2008effect}.

\section{Discussions}
We propose and analyze an age-structured mixed PDE-ODE model of malaria transmission which couples the vector-host epidemiological dynamics with immunity dynamics. Our model tracks the acquisition and loss of immunity due to exposure and waning, and incorporates vaccinations. It also captures the corresponding feedback on the immunity-related epidemiological parameters and characterizes the resulting heterogeneity in age-based immunity distributions.

We prove the well-posedness of the model and analyze the stability of the DFE, the threshold condition of which defines the basic reproduction number, $\mathcal{R}_0$. We then interpret the $\mathcal{R}_0$ as an averaged one-way reproduction number between humans and mosquitoes, collecting the weighted contribution from different infectious stages and age cohorts. Owing to the complexity of our model, numerical bifurcation analysis is required to study the existence and structure of the endemic equilibrium.  

We parametrize and calibrate the model according to a high-transmission setting in sub-Saharan Africa. Our numerical study emphasizes the essential role of immunity dynamics for population level predictions and successfully reproduces various qualitative features of age-infection distributions observed in high-transmission regions. Due to frequent exposure in high-aEIR regions, people develop a high level of anti-disease immunity  over time. As observed from the bifurcation plots, the immunity, in turn, reduces the overall fraction of severe disease. This immunity feedback creates a strong heterogeneity in the age distribution of the immunity profile and infection status. Our numerical results show that severe disease peaks in young children, and more asymptomatic than severe cases are seen among children over 10 years of age.

We also numerically investigate the impact of a vaccine that produces short-lived anti-parasite immunity, motivated by RTS,S, in a simplified setting. Upon completion of the full three-dose series of RTS,S, there is a large drop in severe disease for young children under three years of age. However, the effect may be reversed slightly among older children due to the delay in developing exposure-acquired immunity.

As a proof of concept, our model offers important insights into malaria immunity dynamics; however, we recognize various model assumptions that limit its ability to make quantitative predictions for real-world scenarios. One of the biggest assumptions, for the sake of analytic tractability, is constant population size. 
We calibrated the model to balance the fertility and natural mortality rates. Most developing countries in sub-Saharan Africa have a growing population, and in future work, we will consider the impact this population growth has on the system dynamics and our results.
We also assumed a zero disease-induced mortality rate, but in 2019, an estimated 409,000 people died of malaria, most of which were young children in sub-Saharan Africa~\cite{CDC_data}. Thus, it is of evident interest to explore the impact of disease-induced mortality in future work.

\edit{Although we did not observe multiple endemic equilibria or backward (subcritical) bifurcation in the numerical study presented, these phenomena can occur for complex age-structured models (cf. \cite{inaba2003backward}) and are crucial to understand from the perspective of malaria control~\cite{hadeler1997backward}. Disease-induced mortality may induce backward bifurcations in vector-host malaria models~\cite{chitnis2006bifurcation,gumel2012causes}, including those with age-structure~\cite{richard2022human}, so relaxing the constant population size assumption (as discussed above) further motivates a thorough study of these phenomena. Vaccination has also been shown to induce backward bifurcations in simpler epidemic models~\cite{brauer2004backward,kribs2000simple}, and it would be practically useful to derive conditions for such a bifurcation to occur in our model. Indeed, the ability of vaccination strategies to produce multiple endemic equilibria and backward bifurcations makes it an essential next step to investigate these properties for a more in-depth study of vaccination programs in the present modeling framework.}

Further work could also include a more realistic description of RTS,S vaccine implementation, such as including the boosting dose for two-year-old children to search for an optimal strategy to improve disease outcomes. Additionally, our model provides a framework to study blood-stage vaccines that boost anti-disease immunity ($C_\nu$ variable). Thus, in future work, we will compare the impact of vaccines and vaccination programs that stimulate immunity via different mechanisms. 

\appendix
\section{Well-posedness of the PDE-ODE Model}\label{sec_appendix_wellposed}
\color{black}

We aim to formulate our model as an abstract Cauchy problem of the form
\begin{equation*}\label{eq.abstract_ODE}
\frac{d}{dt}u(t) = \mathcal{A}u(t) + \mathcal{F}(u(t)),\quad t \geq 0, \quad u(0) = u_0,
\end{equation*}
or, more precisely, as an integral equation of the form
\begin{equation}\label{eq.abstract_IE}
u(t) = u_0 + \mathcal{A}\int_0^t u(s)\,ds + \int_0^t \mathcal{F}(u(s))\,ds, \quad t \geq 0.
\end{equation}
When the linear operator $\mathcal{A}$ generates a strongly continuous semigroup and $\mathcal{F}$ is Lipschitz, it is relatively straightforward to establish the desired existence and uniqueness result for an age-structured system~ \cite{inaba1990threshold,webb1985theory}.
Due to the nature of the boundary conditions in our model, $\mathcal{A}$ will not be densely defined and we hence rely on the results of Thieme~\cite{thieme1990semiflows} (see also \cite{cai2013epidemic,martcheva2003progression}). 
Although our calculations will establish an integral solution satisfying \eqref{eq.abstract_IE}, differentiability of the solution can further be established with sufficient smoothness assumptions (see Thieme~\cite[Theorem 3.7]{thieme1990semiflows}). 

To apply the relevant semigroup results, we need to define the operators $\mathcal{A}$ and $\mathcal{F}$ such that:
\begin{enumerate}
\item[(i.)] $\mathcal{A}$ is a closed linear operator on a Banach space $X$, $\lambda-\mathcal{A}$ has a bounded inverse, and for $n\in \mathbb{N}$, 
\[
\| (\lambda-\mathcal{A})^{-n} \| \leq \frac{Z}{(\lambda-\omega)^n}, \quad \mbox{for all }\lambda > \omega,
\]
and $Z$ and $\omega$ positive constants. 
\item[(ii.)] $X_0 := \overline{\mathcal{D}(\mathcal{A})}$ denotes the closure of the domain of $\mathcal{A}$. For some $C$, a closed convex subset of $X$, define
$
C_0 = C \cap X_0$ and take $x_0 \in C_0$.
\item[(iii.)] $\mathcal{F}: C_0 \mapsto X$ is Lipschitz and linearly bounded, i.e. $\| \mathcal{F}(x)\| \leq c(1+\|x\|)$ for some $c>0$.
\end{enumerate}
We consider solutions with a fixed human population structure, i.e.
\begin{align*}
P_H(\alpha,t) = P_H^*(\alpha) := \mu_H^* N_H e^{-M(\alpha)}, \quad
\mu_H^* = \frac{1}{\edit{\int_0^A} e^{-M(\alpha)}d\alpha}, \quad
N_H >0, \text{ for }\alpha \in [0, A).
\end{align*}
We also assume that the mosquito population level begins at its long-run equilibrium level by setting $N_M = g_M/\mu_M$. Replacing $P_H(\alpha,t)$  by $P_H^*(\alpha)$, we work with the normalized system given by \cref{eq.immunity_proportions} for the human disease components and \cref{eq.immunity_proportions2} for the immunity components; the forces of infection in this normalized form are given by
\begin{align*}
\wedge_M(H(t)) &:= b_M(N_M,N_H)\mu_H^*\edit{\int_0^A}\left(\beta_D \widetilde{D}_H(\alpha,t)+ (\beta_A \widetilde{A}_H(\alpha,t)\right)e^{-M(\alpha)}d\alpha,\\
\wedge_H(I_M(t)) &:=b_H(N_M,N_H)\beta_M \,I_M/N_M.
\end{align*}
The vector $H(t) = \left(0,\widetilde{S}_H(t), 0, \widetilde{E}_H(t),\dots,0,\widetilde{C}_{\nu}(t)\right)^T$ is comprised of the human and immune components, with a zero entry for each boundary condition. Thus
\begin{align*}
\frac{dS_M}{dt}&= - \wedge_M(t) S_M +g_M -\mu_M S_M,\\
\frac{dE_M}{dt}&= \wedge_M(t) S_M -\sigma E_M-\mu_M E_M,\\
\frac{dI_M}{dt}&= \sigma E_M - \mu_M I_M,\\
\frac{dH}{dt}&= \widehat{\mathcal{A}}H(t) +\widehat{\mathcal{F}}(S_M;I_M;E_M;H),
\end{align*}
where $\widehat{\mathcal{A}}: ~\mathcal{D}(\widehat{\mathcal{A}}) \mapsto \widehat{X}$ and $\widehat{\mathcal{F}}: \widehat{C_0} \mapsto \widehat{X}$ with
\begin{align*}
\widehat{X} &:= \mathbb{R} \times L^1(0,A)\times \cdots \times \mathbb{R} \times L^1(0,A),\\
\mathcal{D}(\widehat{\mathcal{A}}) &:= \{0\}\times W^{1,1}(0,A)\times \cdots \times  \{0\}\times W^{1,1}(0,A),\\
\widehat{C_0} &:= ~\{0\}\times L^1\big((0,A);[0,1]\big)\times \cdots \times  \{0\}\times L^1\big((0,A);[0,\infty)\big).
\end{align*}
$W^{1,1}(0,A)$ is the space of functions in $L^1(0,A)$ whose weak derivatives are also in $L^1(0,A)$. Let $\vee \in \mathcal{D}(\widehat{\mathcal{A}})$ and define $\widehat{\mathcal{A}}$ by
\begin{align*}
\widehat{\mathcal{A}}\vee = &\Big( -\widetilde{S}_H(0),- \partial_\alpha \widetilde{S}_H- \eta(\alpha)\nu_p(\alpha)\widetilde{S}_H,
 -\widetilde{E}_H(0),-\partial_\alpha \widetilde{E}_H - h \widetilde{E}_H,
-\widetilde{A}_H(0),\\
&\qquad-\partial_\alpha \widetilde{A}_H- r_A\widetilde{A}_H, -\widetilde{D}_H(0),   -\partial_\alpha \widetilde{D}_H- r_D \widetilde{D}_H, -\widetilde{V}_H(0), -\partial_\alpha \widetilde{V}_H-w \widetilde{V}_H, \\ &\qquad -\widetilde{C}_e(0), - \partial_\alpha \widetilde{C}_e - \dfrac{\widetilde{C}_e}{d_e},
 -\widetilde{C}_m(0), -\partial_\alpha \widetilde{C}_m - \dfrac{\widetilde{C}_m}{d_m},
-\widetilde{C}_{\nu}(0),  -\partial_\alpha \widetilde{C}_{\nu} - \dfrac{\widetilde{C}_{\nu}}{d_{\nu}}\Big)^T.
\end{align*}
Similarly, for $\vee \in \widehat{C_0}$ and $(S_M,E_M,I_M) \in [0,N_M]^3$, let $\mathcal{F}(S_M;E_M;I_M;\vee):=$
\begin{align*}
&\Big(1, -\wedge_H\widetilde{S}_H + \phi(\widetilde{C}_H ) r_D\, \widetilde{D}_H + r_A \widetilde{A}_H
+w \widetilde{V}_H, ~0,~ \wedge_H \widetilde{S}_H,\\
&\quad 0,~ (1-\rho(\widetilde{C}_H )) h \widetilde{E}_H - \psi(\widetilde{C}_H) \wedge_H\widetilde{A}_H 
+ (1- \phi(\widetilde{C}_H )) r_D \widetilde{D}_H,\\
&\quad 0,~   \rho(\widetilde{C}_H ) h \widetilde{E}_H + \psi(\widetilde{C}_H) \wedge_H \widetilde{A}_H,~0,~ \eta(\alpha)\,\nu_p(\alpha) \widetilde{S}_H,\\
&\quad 0,~ f(\wedge_H)\left\{ c_S \widetilde{S}_H,+ c_E \widetilde{E}_H + c_A \widetilde{A}_H + c_D \widetilde{D}_H\right\},\\
&\quad m_0\edit{\int_{0}^{A}}g_H (\alpha) e^{-M(\alpha)} \big(c_1\widetilde{C}_e +c_3\widetilde{C}_{\nu}\big)\,d\alpha,~ 0,~
c_\nu \nu_b(0,t),~ c_{\nu} \,\nu_b(\alpha,t) \widetilde{S}_H \Big)^T.
\end{align*}
These choices enforce the boundary conditions of our system. To complete the  setup of the human-mosquito system, let $X = \mathbb{R}\times\mathbb{R}\times\mathbb{R}\times\widehat{X}$ and define
$\mathcal{A}:~ \mathcal{D}(\mathcal{A}) \mapsto X$ by
\[
\mathcal{A}\left(\begin{array}{c}
S_M \\ E_M \\ I_M \\H
\end{array}\right)
= \left(\begin{array}{cccc}
-\mu_M & 0 & 0 & 0 \\ 
0 & -(\sigma +\mu_M) &0 & 0\\
0 & \sigma & -\mu_M & 0 \\ 
0 & 0 & 0 & \widehat{\mathcal{A}}
\end{array}\right)
\left(\begin{array}{c}
S_M \\ E_M \\ I_M \\H
\end{array}\right),
\]
where $\mathcal{D}(\mathcal{A}) = \mathbb{R}\times\mathbb{R}\times\mathbb{R}\times\mathcal{D}(\widehat{\mathcal{A}})$. Hence
\begin{align*}
\overline{\mathcal{D}(\mathcal{A})} &= \mathbb{R}\times\mathbb{R}\times\mathbb{R}\times \{0\}\times L^1(0,A)\times \cdots \times  \{0\}\times L^1(0,A) =: X_0 \subset X,
\end{align*}
and we may choose
\[
C = [0,N_M]\times \cdots \times [0,N_M] \times \{0\}\times L^1\big((0,A);[0,1]\big)\times \cdots \times  \{0\}\times L^1(0,A),
\]
which is a closed, convex subset of $X$ and gives $C=C_0$. The human disease components are in $L^1\big((0,A);[0,1]\big)$, while the human immunity components are in $L^1(0,A)$. Define $\mathcal{F}:~C_0\mapsto X$ by
\[
\mathcal{F}\left(\begin{array}{c}
S_M \\ E_M \\ I_M \\H
\end{array}\right)
= \left(\begin{array}{c}
\wedge_M S_M + g_M\\ 
\wedge_M S_M\\
0 \\ 
\widehat{\mathcal{F}}(S_M;E_M;I_M;H)
\end{array}\right).
\]
In the following, the norm on each space is always the natural one, for example,
\[
\|\vee\|_{\widehat{X}} = |\vee_1| + \|\vee_2\|_{L^1}+ |\vee_3| + \|\vee_4\|_{L^1}+\cdots .
\]
To show that $\lambda-\mathcal{A}$ has a bounded inverse and verify condition (i.), solve
$
(\lambda-\mathcal{A})\vee = f$ for $ \vee\in\mathcal{D}(\mathcal{A})$ and $f\in X
$. To this end, let
$
\vee = (S_M,E_M,I_M,0,\widetilde{S}_H,0,\widetilde{E}_H, \dots, 0,\widetilde{C}_{\nu})^T
$ 
and
$
f = (f_{S_M},f_{E_M},f_{I_M},f_{S_1},f_{S_2},\dots,f_{{\nu}_1},f_{{\nu}_2})^T
$, and write
\[
(\lambda-\mathcal{A})\vee = \left(\begin{array}{cccc}
\lambda+\mu_M & 0 & 0 & 0 \\ 
0 & \lambda+\sigma +\mu_M &0 & 0\\
0 & -\sigma & \lambda+\mu_M & 0 \\ 
0 & 0 & 0 & \lambda-\widehat{\mathcal{A}}
\end{array}\right)
\left(\begin{array}{c}
S_M \\ E_M \\ I_M \\(0, \dots,\widetilde{C}_{\nu})^T
\end{array}\right)=f.
\]
Solving the equation above yields
$
S_M = f_1/(\lambda+\mu_M)
$, 
$E_M = f_2/(\lambda+\sigma+\mu_M),$ and $
I_M =  f_3/(\lambda+\mu_M)+\sigma f_2/\left((\lambda+\mu_M)(\lambda+\sigma+\mu_M)\right)$. The solutions for the PDE components are largely repetitive; for example, solving $\partial_\alpha \widetilde{S}_H + (\lambda + \eta(\alpha)\nu_p(\alpha))\widetilde{S}_H = f_{S_2}$ gives 
\[
\widetilde{S}_H(\alpha) =\exp\left( -\int_0^{\alpha}\big(\lambda + \eta(a)\nu_p(a)\big)da\right)\left(f_{S_1}+\int_0^{\alpha}e^{\int_0^a(\lambda  + \eta(s)\nu_p(s))ds}f_{S_2}(a)da\right).
\]
Next estimate the norm of each component to plug into the estimate on $\|(\lambda-\mathcal{A})^{-1}f\|$. For the most part, these are straightforward and follow similarly. For example,
\begin{align*}
\| \widetilde{E}_H\|_{L^1} & \leq |f_{E_1}| \left(\left. \frac{-e^{-(\lambda+h)\alpha}}{\lambda+h}\right|_0^A + \edit{\int_0^A}\int_a^A e^{-(\lambda+h)(\alpha-a)}\left|f_{E_2}(a)\right|d\alpha\,da \right)\\
&\leq \frac{\left|f_{E_1}\right|}{\lambda+h} +\frac{\|f_{E_2}\|_{L^1}}{\lambda+h}, \quad \mbox{for }\lambda>-h.
\end{align*}
The estimate for the susceptible components requires use of the assumption (H5), which guarantees that $\text{essinf}_{\alpha\in[0,A)} \eta(\alpha)\nu_p(\alpha) =: \underline{\eta}>0$. Hence 
\[
\|\widetilde{S}_H\|_{L^1} \leq \frac{\left|f_{S_1}\right|}{\lambda+\underline{\eta}} +\frac{\|f_{S_2}\|_{L^1}}{\lambda+\underline{\eta}},\quad \mbox{for }\lambda > - \underline{\eta}.
\]
Choose $\epsilon^*:=\min\{\mu_M,\underline{\eta},h,r_A,r_D,w,1/d_e,1/d_m,1/d_{\nu}\}$, and let $\lambda > -\epsilon^*$. Then,  
$
\|(\lambda-\mathcal{A})^{-1}f\| \leq \|f\|/(\lambda + \epsilon^*),
$
and hence
\[
\|(\lambda-\mathcal{A})^{-1}\| \leq \frac{1}{(\lambda + \epsilon^*)^n},\quad \mbox{for all } n \in \mathbb{N}.
\]
\noindent It is straightforward to show that $\mathcal{F}$ obeys condition (iii.). The Lipschitz condition on $\mathcal{F}$ follows from the Lipschitz conditions on $f$, $\phi$, $\rho$ and $\psi$, and the linear boundedness estimate follows from the boundedness of the human/mosquito components and the  coefficients of the system.

The following result is a consequence of Theorems 2.3 and 3.2 from Thieme~\cite{thieme1990semiflows}.
\begin{theorem}
If the following conditions hold:
\begin{enumerate}
\item[(a.)] $\lambda(\lambda-\mathcal{A})^{-1}$ maps $C$ to $C$ for $\lambda$ sufficiently large,
\item[(b.)] $\frac{1}{h}\text{dist}\left(\vee+h \mathcal{F}(\vee); C\right) \rightarrow 0$ as $h\downarrow 0$ for all $\vee \in C_0$,
\end{enumerate}
then there exists a unique continuous solution to \eqref{eq.abstract_IE} with values in $C_0$. Moreover, the map $U:\mathbb{R}_+ \times C_0 \mapsto C_0$ defined by  $U(t,u_0) := u(t)$ is a continuous semiflow and satisfies an exponential Lipschitz condition, i.e.,
\[
\|U(t,u_0) - U(t,u_1) \| \leq J\,e^{ct}\|u_0 -u_1 \|, \quad J\geq 1,\quad c\in\mathbb{R}, \quad u_0,u_1\in C_0, \quad t \geq 0.
\]
\end{theorem}
\begin{proof}
To verify that $(a.)$ holds, note that $(\lambda-\mathcal{A})^{-1}$ clearly preserves positivity and integrability, and thus so does $\lambda(\lambda-\mathcal{A})^{-1}$ for any $\lambda>0$. Checking that upper bounds on the components in $C$ are not violated requires solving $\lambda^{-1}(\lambda-\mathcal{A})\vee = f$ with $\vee, f \in C$. For example, for the first mosquito component, we obtain
$
\lambda f_1/(\lambda+\mu_M)
$,
which is in $[0,N_M]$ since $\lambda/(\lambda +\mu_M) < 1$ for $\lambda>0$ and $f_1 \in [0,N_M]$. The other mosquito components work similarly for $\lambda>0$. The human disease components all follow a similar pattern. For example, computing $\lambda(\lambda-\mathcal{A})^{-1}\widetilde{E}_H$ yields
\begin{align*}
\lambda f_{E_1} e^{-(\lambda+h)\alpha}&+ \lambda\int_0^{\alpha}e^{-(\lambda+h)(\alpha-a)}f_{E_2}(a)\,da \leq \frac{\lambda}{\lambda+h}+{\lambda f_{E_1}e^{-(\lambda+h)\alpha}}.
\end{align*}
The first term on the right-hand side is less than or equal to 1 for $\lambda>0$ and the second term tends to zero as $\lambda \to \infty$, so $\lambda(\lambda-\mathcal{A})^{-1}\widetilde{E}_H < 1$ for $\lambda$ sufficiently large.

To check condition $(b.)$, form the operator
$
\widetilde{\mathcal{F}}(\vee) := \mathcal{F}(\vee) + \gamma \vee, $ for $\vee\in C_0$, and choose $\gamma>0$ sufficiently large that
\begin{equation}\label{eq.admissible}
\widetilde{\mathcal{F}}(\vee): C_0 \mapsto X_+.
\end{equation}
If such a $\gamma$ exists, then, for any $\vee \in C_0$,
\begin{align*}
\frac{1}{h}\text{dist}\left(\vee+h \mathcal{F}(\vee); C\right) &= \frac{1}{h}\text{dist}\left(\vee-\gamma h\vee + h \widetilde{\mathcal{F}}(\vee); C\right) =0 \quad \text{for $h$ sufficiently small}.
\end{align*}
The second equality holds because, for a fixed $\gamma$ guaranteeing \eqref{eq.admissible}, both $\vee-\gamma h \vee\in C$ and $h\widetilde{\mathcal{F}}(\vee)\in C$ for $h$ sufficiently small. The latter inclusion follows because $\widetilde{\mathcal{F}}(\vee)\in X_+$ so  $h \widetilde{\mathcal{F}}(\vee)$ will be non-negative but some components may exceed the pointwise bounds required by $C$; scaling by $h$ sufficiently small remedies this and thus $(b.)$ holds.

For $\vee\in C_0$, $\mathcal{F}(\vee)$ is bounded in each human disease and mosquito component so finding a $\gamma$ guaranteeing \eqref{eq.admissible} is straightforward. For example, 
$$
\mathcal{F}(S_M) + \gamma S_M = -S_M b_M \mu_H^* \edit{\int_0^A} \left( \beta_D \widetilde{D}_H + \beta_A \widetilde{A}_H\right) e^{-M(\alpha)}d\alpha + g_M + \gamma S_M \geq 0,
$$
if $\gamma > b_M(\beta_D+\beta_A)$. 
\end{proof}
\color{black}
\section{Calculations for the Proof of \texorpdfstring{\cref{thm:R0threshold}}{\ref{thm:R0threshold}}}\label{sec:appendixR0}
To see that $|\zeta(p)| \leq \zeta\left( \text{Re}(p) \right)$ for $p \in \mathbb{C}$, estimate as follows:
\begin{align*}
\left|\mathcal{E}(\alpha,p)\right| &= \int_0^\alpha \left| e^{-(h+p)(\alpha-a)}\right| \theta(a)\,da = \int_0^\alpha  e^{-(h+\text{Re}(p))(\alpha-a)} \left|\edit{ e^{-i\,\text{Im}(p)(\alpha-a)}}\right| \theta(a)\,da \\ &\leq \int_0^\alpha  e^{-(h+\text{Re}(p))(\alpha-a)} \theta(a)\,da \leq \mathcal{E}(\alpha,\text{Re}(p)).
\end{align*}
It follows from the inequality above and analogous reasoning that 
\[
\left|\mathcal{D}(\alpha,p)\right| \leq \mathcal{D}(\alpha,\text{Re}(p)), \quad \left|\mathcal{A}(\alpha,p)\right| \leq \mathcal{A}(\alpha,\text{Re}(p)).
\]
Therefore $|\zeta(p)| \leq \zeta\left( \text{Re}(p) \right)$, as claimed.

To see that $\zeta$ in non-increasing for nonnegative real arguments, observe that
\[
\partial_p \mathcal{E}(\alpha,p) = \int_0^\alpha (a-\alpha) e^{-(h+p)(\alpha-a)}\theta(a)\,da  \leq 0, \quad \mbox{for } \alpha \geq 0, \,\, p \geq 0.
\]
Similarly, 
\begin{align*}
\partial_p \mathcal{D}(\alpha,p) &= h \int_0^\alpha e^{-(r_D+p)(\alpha-a)} \rho(\widetilde{C}_{H}^*(a)) \frac{\partial}{\partial p} \mathcal{E}(a,p)\,da \\ &\,\,+ h \int_0^\alpha (a-\alpha) e^{-(r_D+p)(\alpha-a)} \rho(\widetilde{C}_{H}^*(a)) \mathcal{E}(a,p)\,da \leq 0, \quad\mbox{for } \alpha \geq 0, \,\, p \geq 0,
\end{align*}
since both terms on the right-hand side are less than or equal to zero. An analogous calculation, relying on the inequalities above, shows that $\partial_p \mathcal{A}(\alpha,p) \leq 0$ for $(\alpha,p) \in \mathbb{R}_+^2$. Therefore $p \mapsto \zeta(p)$ is nonincreasing for $p \in \mathbb{R}_+$, as claimed.  

Finally, we claimed that 
\[
\lim_{p \to \infty}\mathcal{D}(\alpha,p) = \lim_{p \to \infty}\mathcal{A}(\alpha,p) = 0\quad (\text{uniformly in $\alpha$}).
\]
It is straightforward to see that $\mathcal{E}(\alpha,p) \leq 1/(h+p)$ and hence
\[
\mathcal{D}(\alpha,p) \leq \frac{h}{h+p} \int_0^\alpha e^{-(r_D+p)(\alpha-a)}\,da \leq \frac{h}{(h+p)(r_D+p)},
\]
immediately confirming the first part of the claim. Using these estimates and performing similar upper bounding on $\mathcal{A}(\alpha,p)$ shows that
\[
\mathcal{A}(\alpha,p) \leq  \frac{h}{(h+p)(r_A+p)} + \frac{h\,r_D}{(h+p)(r_D+p)(r_A+p)},
\]
and hence the second part of the claim follows immediately.

\section{Supplementary Figures}
\begin{figure}[htbp]
    \centering
    \includegraphics[width=0.9\textwidth]{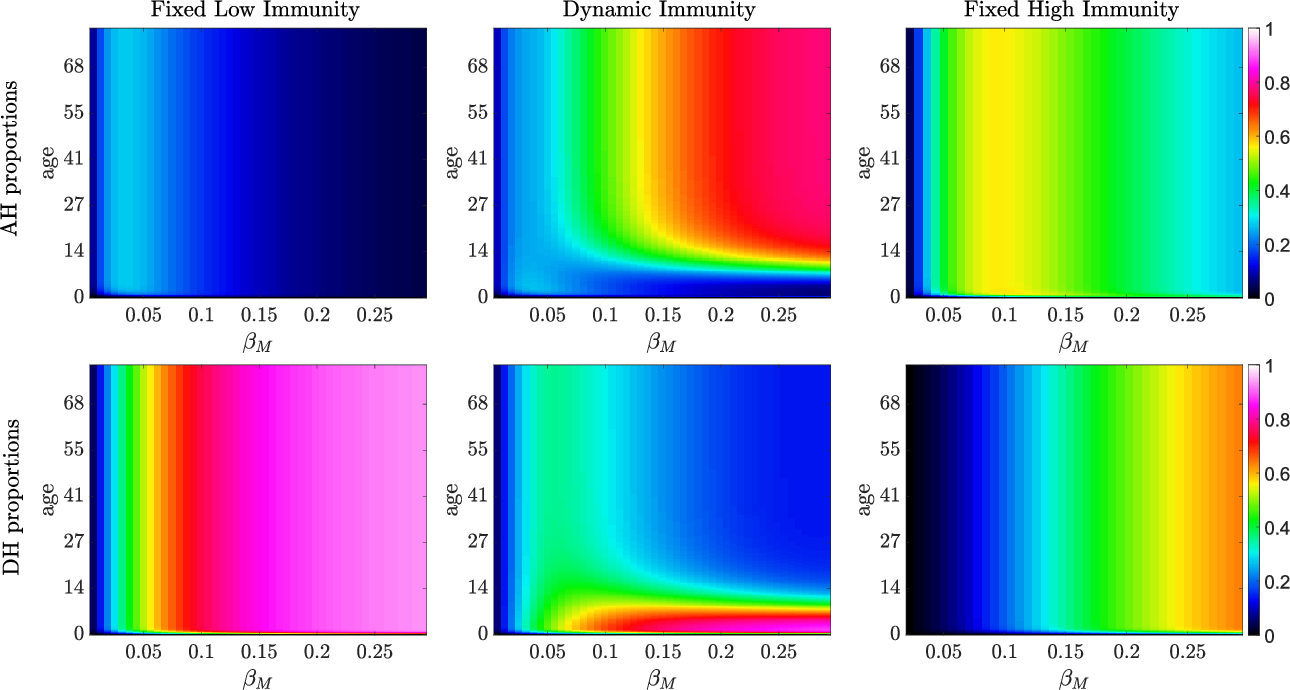}
    \caption{Plot showing the underlying age structure of the endemic equilibrium in the model with dynamic immunity (center) versus fixed low (left) and high (right) immunity scenarios.}
    \label{fig:supp_fig_1}
\end{figure}

\Cref{fig:supp_fig_1} illustrates the age structure of the endemic equilibrium and provides additional detail to the bifurcation diagrams in \cref{fig:comp_bifur}.

\Cref{fig:supp_fig_2} below shows the same bifurcation diagrams as in \cref{fig:comp_bifur} but with $\mathcal{R}_0(\beta_M)$ on the x-axis.
\begin{figure}[htbp]
    \centering
    \includegraphics[width=0.325\textwidth]{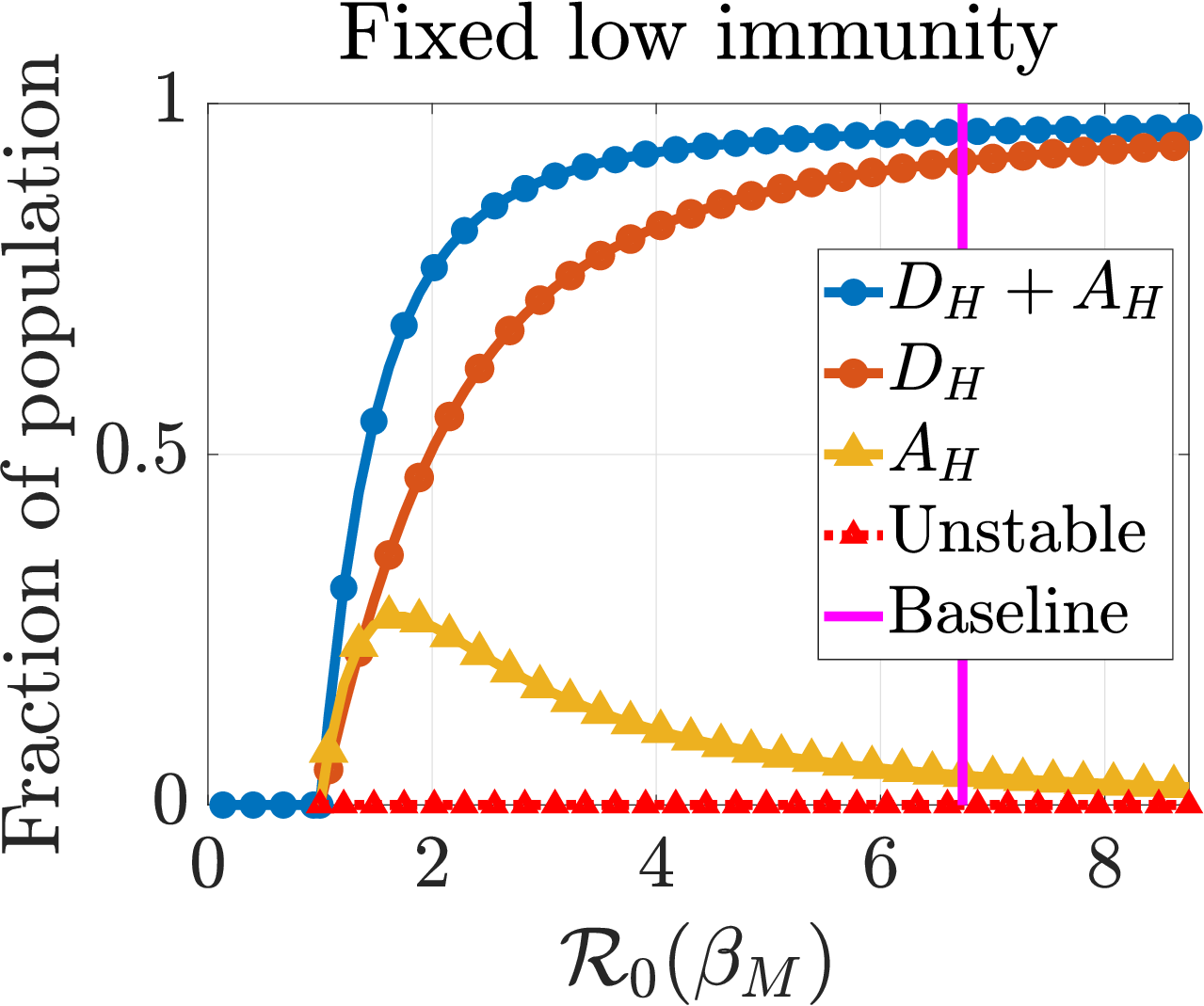}\hfill\includegraphics[width=0.325\textwidth]{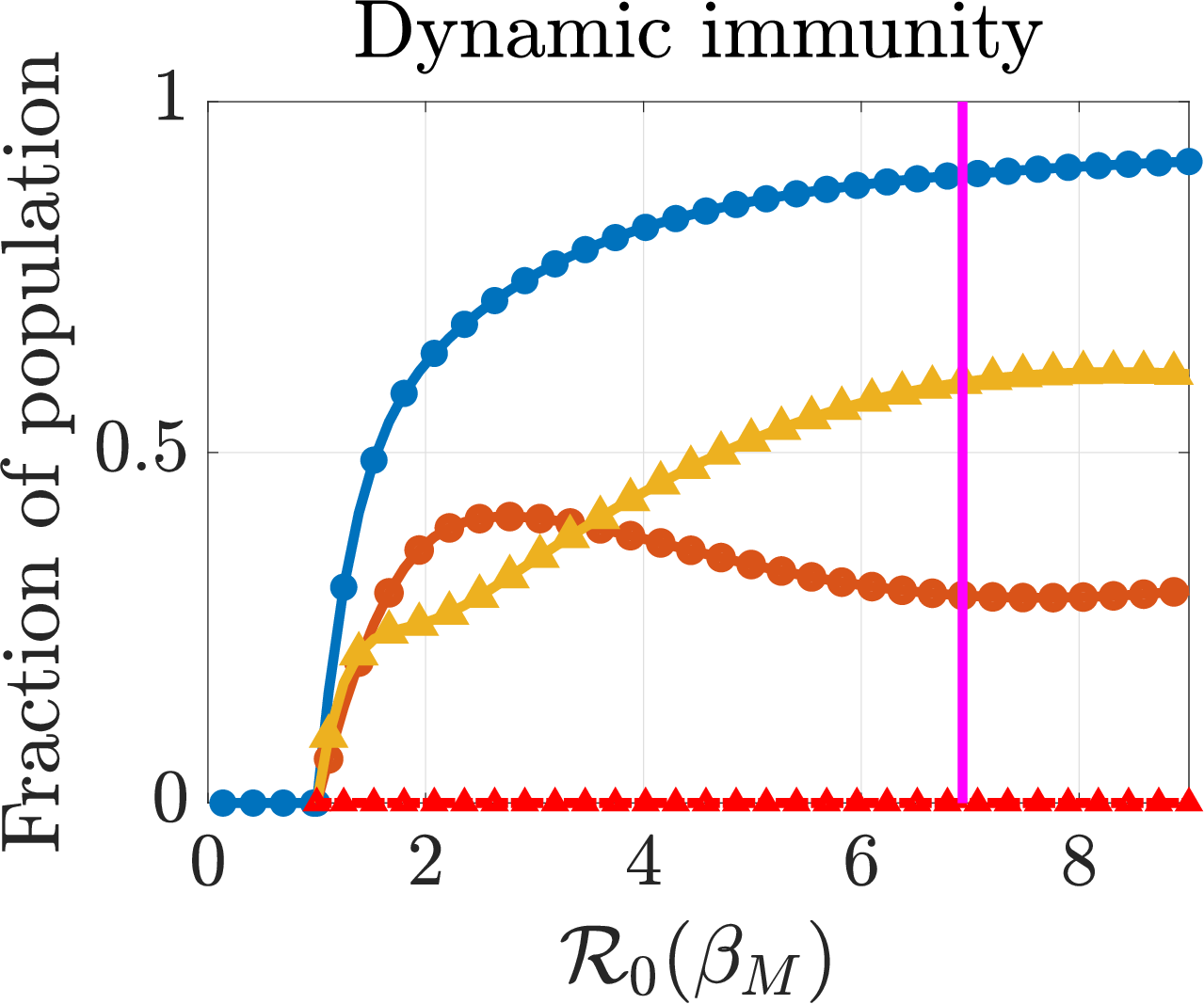}\hfill\includegraphics[width=0.325\textwidth]{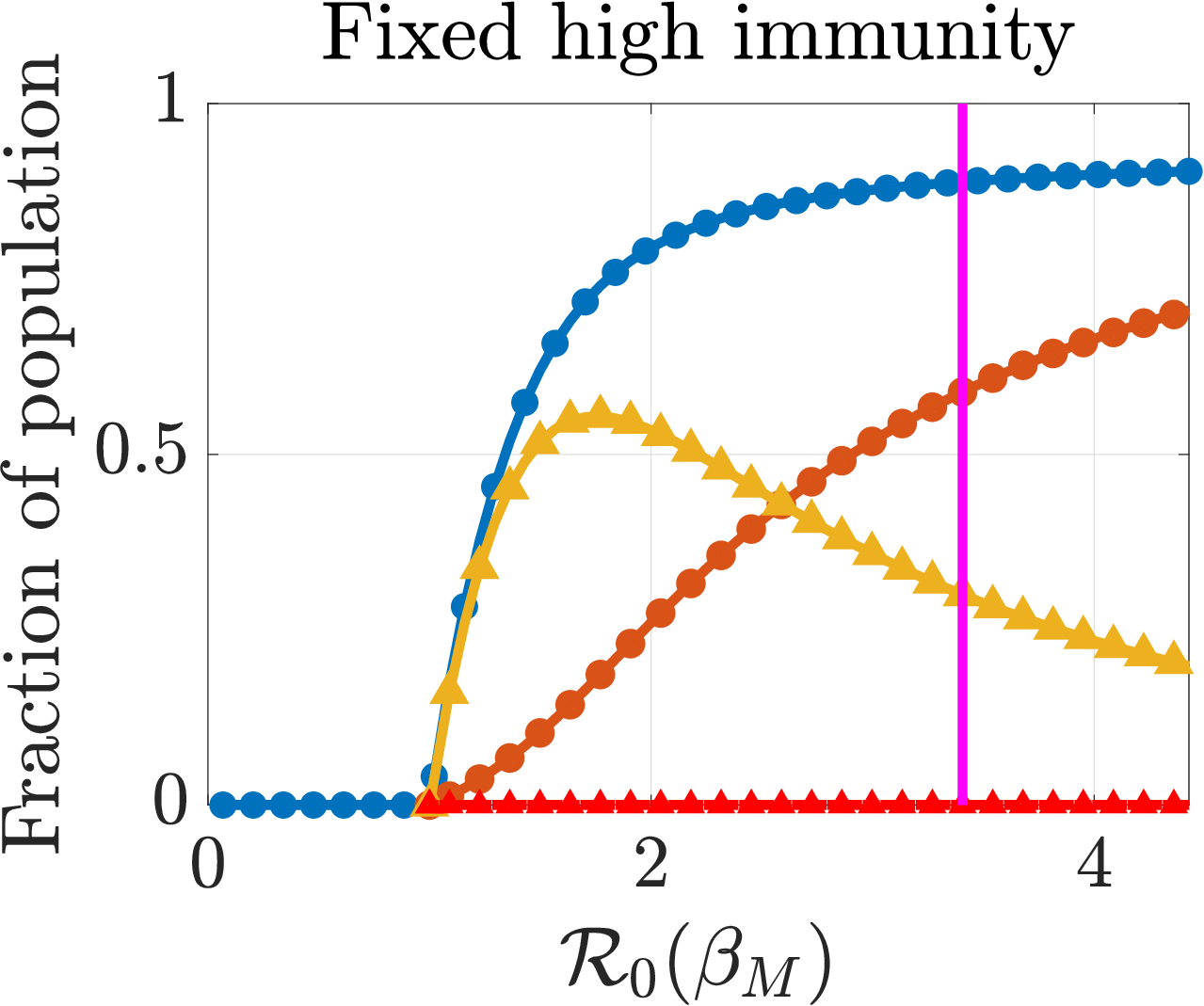}
    \caption{Bifurcation analysis with $\mathcal{R}_0(\beta_M)$ on the x-axis.}
    \label{fig:supp_fig_2}
\end{figure}

\begin{figure}[htbp]
    \centering
    \includegraphics[width=0.5\textwidth]{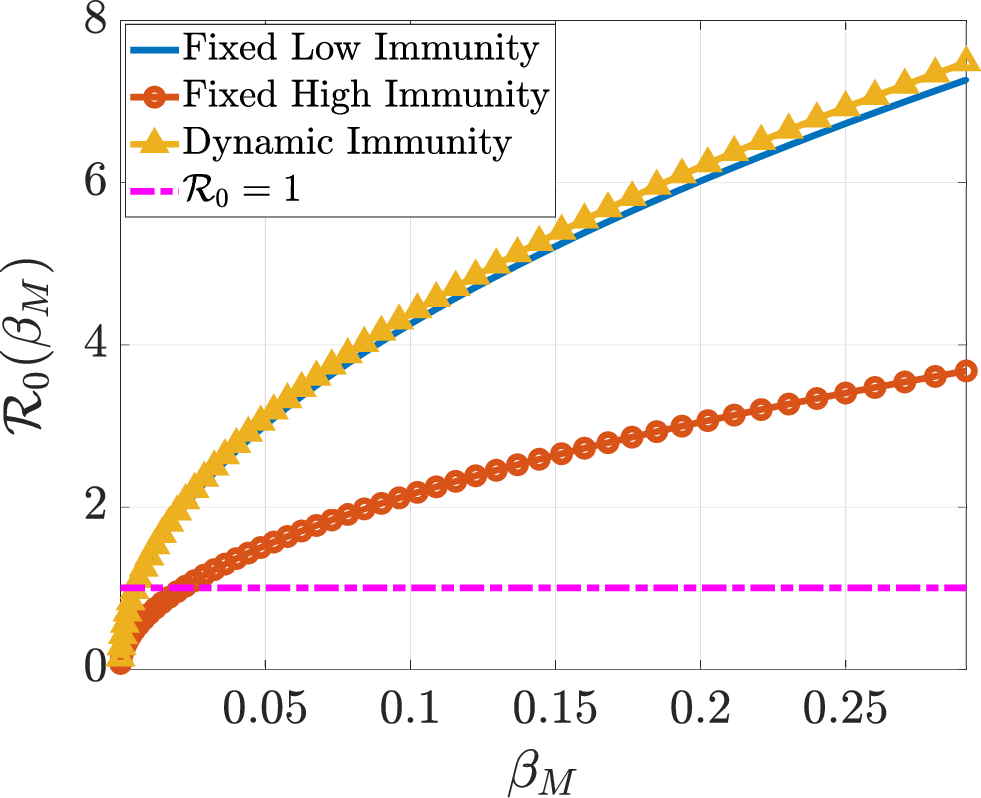}
    \caption{$\mathcal{R}_0$ as a function of $\beta_M$ for each of the three immunity scenarios studied in the paper.} \label{fig:supp_fig_3}
\end{figure}

\Cref{fig:supp_fig_3} shows $\mathcal{R}_0$ as a function of $\beta_M$ for each of the three scenarios presented in the paper.

\newpage
\bibliographystyle{siamplain}
\bibliography{References}
\end{document}